\documentclass[aps,prb]{revtex4-2}

\usepackage{graphicx}
\usepackage[utf8]{inputenc}
\usepackage{amsmath}
\usepackage{comment}
\usepackage{bm}
\usepackage{dsfont}
\usepackage{amsfonts}
\usepackage{xcolor}
\usepackage{booktabs}
\usepackage{multirow}
\usepackage[normalem]{ulem}

\usepackage[mathlines]{lineno}
\usepackage{etoolbox}
\usepackage{hyperref}

\newcommand{\bluecolor}[1]{\textcolor{black}{#1}}

\begin{document}

\title{\bluecolor{Two topological phases in exchange alternating spin-1 nanographene chains}}

\author{João Henriques$^{1,2}$\footnote{joao.henriques@inl.int}, 
Yelko del Castillo$^{1,3}$,
Ricardo Segundo$^{1,4}$, 
Jan Phillips$^1$
Joaquín Fern\'{a}ndez-Rossier$^1$\footnote{On permanent leave from Departamento de F\'{i}sica, Universidad de Alicante, 03690 San Vicente del Raspeig, Spain}$^,$ }

\affiliation{$^1$International Iberian Nanotechnology Laboratory (INL), Av. Mestre Jos\'{e} Veiga, 4715-330 Braga, Portugal }
\affiliation{$^2$University of Santiago de Compostela, 15782, Santiago de Compostela, Spain.}
\affiliation{$^3$Centro de F\'{i}sica das Universidades do Minho e do Porto, Universidade do Minho, Campus de Gualtar, 4710-057 Braga, Portugal }
\affiliation{$^4$NOVA School of Science and Technology (FCT NOVA), 2829-516, Caparica, Portugal }

\date{\today}

\begin{abstract}
Magnetic nanographenes are emerging as versatile building blocks for artificial spin lattices, enabling the exploration of flagship one-dimensional quantum-magnetism models with unprecedented control. The spin-1 Heisenberg model, including \bluecolor{bilinear and biquadratic} exchange, was first realized using [3]-triangulenes, revealing the Haldane phase. More recently, Clar’s goblets enabled the spin-1/2 Heisenberg model with exchange alternation, uncovering additional topological phases. 
\bluecolor{Here we show that spin-1 nanographenes can be used to explore bond-alternating chains both in the Haldane phase and beyond it, in a dimerized phase with emergent edge spin-1}. We  use density matrix renormalization group (DMRG) to analyze how \bluecolor{biquadratic} exchange, which is known to be large in spin-1 nanographenes, determines the phase transition boundary. 
Combining multiconfigurational and first-principles calculations, we identify two realistic candidates to realize these \bluecolor{two different } phases: the recently synthesized extended Clar’s goblet and a passivated [4]-triangulene. 
We demonstrate how to distinguish these phases experimentally using inelastic electron tunneling spectroscopy, paving the way for their observation.
\end{abstract}

\maketitle

\section{Introduction}
\label{sec:intro}
Traditionally, the discovery of materials has preceded their exploitation for applications. However, our understanding of the non-trivial properties of certain quantum Hamiltonians \cite{kitaev2001unpaired,kitaev2006anyons,affleck1988valence, affleck1987rigorous} has led to an attempt to reverse the process, and look for  materials and nanostructures that realize a given model \cite{winter2017models}. In this context, the fabrication of artificial lattices made with a variety of building blocks, such as quantum dots \cite{hensgens2017quantum}, superconducting circuits \cite{houck2012chip}, adatoms \cite{khajetoorians2019creating} and cold atoms \cite{mazurenko2017cold} is being heavily explored.

Nanographenes are another example of building blocks that can be used to artificially realize quantum models. The prediction that nanographenes host open-shell ground states is an old one \cite{longuet1950some, ovchinnikov1978multiplicity, rajca1994organic, fernandez2007magnetism, yazyev2010emergence}, and their spinful ground state can be anticipated using the Ovchinnikov rule \cite{ovchinnikov1978multiplicity} which in turn can be related to Lieb's theorem \cite{lieb1989two}. Historically, the high reactivity of these systems limited their experimental investigation to ensemble-averaged techniques. However, the advent of on-surface synthesis \cite{pavlivcek2017synthesis, mishra2019synthesis, su2020triangulenes, li2020uncovering}, combined with high-resolution scanning tunneling microscopy (STM), has enabled direct access to the magnetic properties of individual open-shell nanographenes, such as triangulenes \cite{turco2023observation, pavlivcek2017synthesis, mishra2019synthesis, su2019atomically, su2020triangulenes, mishra2021synthesis}, Olympicene \cite{mistry2015synthesis} and Clar's goblet \cite{clar1972circobiphenyl,mishra2020topological}, to name a few. Therefore, combining the synthetic control of organic chemistry with the atomic control offered by STM, open-shell nanographenes offer a bottom-up route to realizing strongly correlated carbon-based quantum systems \cite{song2024highly}.

\begin{figure*}
    \centering
    \includegraphics[width = \linewidth]{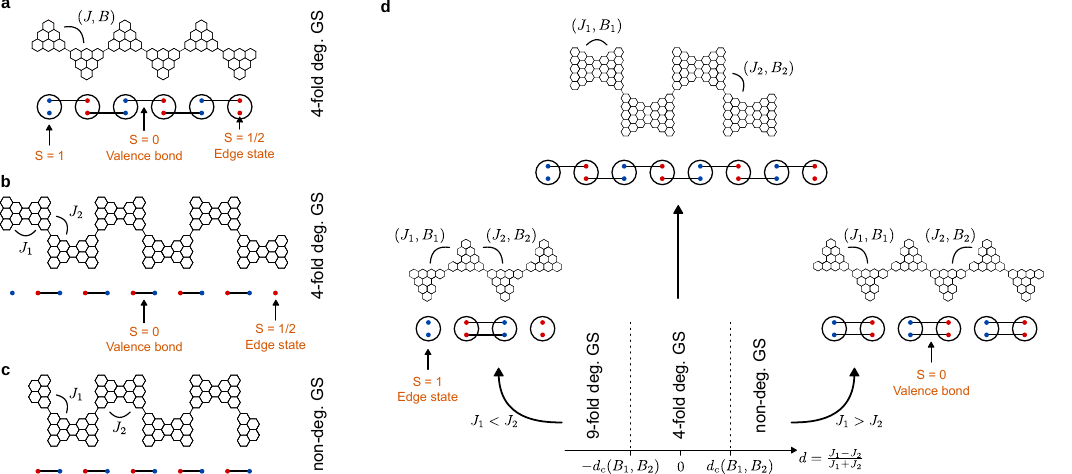}
    \caption{\textbf{Nanographene spin chains.} \textbf{a}, Chain of [3]-triangulenes which realizes the BLBQ Hamiltonian. A depiction of the ground state wave function in the Haldane phase is also shown \cite{mishra2021observation}; \textbf{b}, \textbf{c}, $S=1/2$ BAH Hamiltonian, with $J_1 < J_2$ and $J_1 > J_2$ respectively, realized with Olympicenes \cite{zhao2024tunable}, and the respective depictions of the ground state wave function. \textbf{d}, Phase diagram of the $S=1$ BAH Hamiltonian with bilinear ($J_1$, $J_2$) and biquadratic ($B_1$, $B_2$) exchange couplings. The nanographene chains which realize each phase are depicted, as well as their ground state wave functions.}
    \label{fig:Spin Chains}
\end{figure*}

Recently, nanographenes have been used to study prominent 1D quantum magnetic systems \cite{choi2019colloquium}, starting in Ref. \cite{mishra2021observation} with  $S=1$ spin chains obtained with the [3]-triangulene as the primary unit \cite{martinez2023electrically, huang2026superconducting}. There, it was shown that these spin chains realize  a symmetry protected topological phase, the Haldane phase, with a gap in the excitation spectrum for periodic boundary conditions and a fourfold degenerate ground state for open boundary conditions in the thermodynamic limit. The fourfold manifold is associated with the emergence of fractional $S=1/2$ edge spins. The system was modeled with the bilinear biquadratic (BLBQ) spin Hamiltonian:
\begin{align}
    H^{S=1}_\textrm{BLBQ} = \sum_{i=0} J \boldsymbol{S}_i \cdot \boldsymbol{S}_{i+1} + B (\boldsymbol{S}_i \cdot \boldsymbol{S}_{i+1})^2 \label{eq: BLBQ}
\end{align}
with $J$ being the bilinear exchange, and $B$ quantifying the strength of the biquadratic exchange. This model contains as particular cases both the Heisenberg, $\beta=B/J=0$,  and the AKLT \cite{affleck1988valence}, $\beta=1/3$, Hamiltonians whose ground state wave function admits a simple pictorial representation. The latter, can be written as a sum of spin projectors, and its ground state wave function can be easily depicted, as we show in Fig. \ref{fig:Spin Chains}a. \bluecolor{In a subsequent work \cite{henriques2023anatomy}, a more complete spin model was derived from perturbation theory, for the same system, including second neighbor exchange and three spin interactions. These new terms, however, were found to be smaller than $J$ and $B$, and did not break the system's topological phase.}

Following this breakthrough, the bond alternating Heisenberg spin-1/2 Hamiltonian \cite{hida1992crossover}, was realized with atomic spins on surfaces \cite{wang2024construction} and, more relevant for this work, by covalently linking Olympicenes (which form Clar's goblets) \cite{zhao2024tunable}, as depicted in Fig. \ref{fig:Spin Chains}b and c. In this case, it was found that the system can be described by the bond alternating Heisenberg (BAH) Hamiltonian:
\begin{align}
    H^{S=1/2}_\textrm{BAH} = \sum_{i=0} J_1 \boldsymbol{S}_{2i} \cdot \boldsymbol{S}_{2i+1} + J_2 \boldsymbol{S}_{2i+1} \cdot \boldsymbol{S}_{2i+2} \label{eq: BAH S=1/2}
\end{align}
where $J_1$ and $J_2$ are the \bluecolor{bilinear} exchange interactions in consecutive bonds, which define a dimerization parameter
\begin{align}
    d = \frac{J_1 - J_2}{J_1 + J_2}
\end{align}
This exchange alternation comes about mainly from the spatially inhomogeneous distribution of the zero energy orbital of an Olympicene, which leads to different strengths of kinetic driven exchange \cite{jacob22} depending on whether these molecules are linked tip to tip, or side to side. 

By controlling the type of exchange at the edges of the chain, i.e. terminating it with the stronger ($d>0$) or weaker exchange ($d<0$), it is possible to create spin chains which realize two different topological phases: the odd, or the even Haldane phase, respectively \cite{hida1992crossover}. These two phases are distinguished by the degeneracy of the ground state of infinite chains, which vanishes in the former, and is fourfold degenerate in the latter (a system which is adiabatically connected to the [3]-triangulene chain \cite{mishra2021observation}). The lack of degeneracy in the first case can be explained by a valence-bond solid picture of the ground state, while in the second we have emergent spin-1/2 appearing at the edges, in a similar fashion to what was found in the BLBQ Hamiltonian. Pictorial representations of these ground states are shown in Fig. \ref{fig:Spin Chains}b and c. At the boundary between these two distinct topological phases there is a gapless critical point when $d=0$, a system studied experimentally with various types of nanographenes \cite{zhao2024gapless, yuan2025fractional,su2024fabrication,sun2025surface}.

Given this context, it is natural to inquire whether a  topological  quantum phase transition can also occur with bond-alternating $S=1$ chains made with nanographenes. Such a phase transition was studied \cite{kato1994numerical, totsuka1995isotropic,kitazawa1996phase} for the $S=1$ version of the Hamiltonian in equation (\ref{eq: BAH S=1/2}), ignoring thereby \bluecolor{biquadratic} exchange. Using DMRG, it was found that when the dimerization  parameter is in the region $-0.25 \pm 0.01 < d_c < 0.25 \pm 0.01$ \cite{kato1994numerical}, the system remains in the Haldane phase, with a fourfold degenerate ground state in the thermodynamic limit under open boundary conditions. For bigger absolute values of dimerization, however, two distinct topological phases appear, with different ground state degeneracies: i) ending the chain with a strong bond leads to a singlet-dimer phase with a non-degenerate ground state (self-evident in the limit $d = 1$ since the ground state wave function would simply be a product of singlets); ii) ending the chain with a weak bond produces a ninefold degenerate ground state in the thermodynamic limit due to emergent spin-1 at the chain terminations (the natural extension to what was found for the spin-$1/2$ BAH chain of equation \ref{eq: BAH S=1/2}). Schematic representations of the possible ground states are given in Fig. \ref{fig:Spin Chains}d.

Given that \bluecolor{biquadratic} exchange plays a significant role in spin-1 nanographenes \cite{mishra2021observation,henriques2023anatomy}, the natural model for bond-alternating spin-1 nanographene chains is given by:
\begin{align}
    H^{S=1}_\textrm{BAH} & = \sum_{i=0} J_1 \boldsymbol{S}_{2i} \cdot \boldsymbol{S}_{2i+1} + J_2 \boldsymbol{S}_{2i+1} \cdot \boldsymbol{S}_{2i+2}
    + \sum_{i=0} B_1 (\boldsymbol{S}_{2i} \cdot \boldsymbol{S}_{2i+1})^2 + B_2(\boldsymbol{S}_{2i+1} \cdot \boldsymbol{S}_{2i+2})^2 \label{Eq: S=1 BAH}
\end{align}
with $J_1$ and $J_2$ characterizing the strength of bilinear exchange, and $B_1$ and $B_2$ the biquadratic exchange couplings in consecutive bonds. 
The critical dimerization, $d_c$, for the existence of a phase transition will now be a function of these additional interactions. This can be seen, for instance, when \bluecolor{$\beta_1 = B_1 / J_1 = 1/3$ and $\beta_2 = B_2 / J_2 = 1/3$}: without dimerization ($J_1 = J_2)$, the system is in the Haldane phase (the AKLT limit \cite{affleck1988valence}); in the fully dimerized case, the Hamiltonian remains a sum of projectors (half of which vanish) and still admits the AKLT state as its ground state. Thus, while for $\beta_1 = \beta_2 = 0$ the transition occurs at $|d_c| = 0.25 \pm 0.01$ \cite{kato1994numerical}, for $\beta_1 = \beta_2 = 1/3$ no transition occurs, evidencing the role of biquadratic exchange on $d_c$. The study of the dependence of the boundary of the phase transition as a function of these interaction has previously been considered for the particular case $\beta_1 = \beta_2$, but a thorough study on how this phase boundary depends on $(\beta_1,\beta_2)$ has not been done yet \cite{totsuka1995isotropic}. \bluecolor{Similarly to the Hamiltonian of equation \ref{eq: BLBQ}, terms like second neighbor or three spin terms could have been considered here. However, for the sake of simplicity, we opt to not consider them in the present analysis; in the appendix \ref{sec:additional} we will show quantitatively that, in the systems we consider,  the impact of these terms is small  and they can safely be ignored.}

In this work we first undertake a systematic study of the influence of \bluecolor{biquadratic} exchange interactions on the critical dimerization parameter $d_c$ over the $(\beta_1,\beta_2)$ plane. Afterwards, inspired by the recent synthesis of an extended Clar's goblet \cite{mishra2025synthesis, li2025designer}, we propose feasible nanographene molecules which might realize this model Hamiltonian, and simulate how inelastic electron tunneling spectroscopy can be used to distinguish the different topological phases. Finally, we discuss the experimental feasibility of the dimerized phase taking into account the disorder in the exchange couplings likely to be present in the synthesized chains. 

\section{Critical dimerization map}

\bluecolor{In order to obtain $d_c$ for a given ($\beta_1, \beta_2$) pair, we solve the Hamiltonian in equation (\ref{Eq: S=1 BAH}) with DMRG  \cite{fishman_itensor_2022}, and study the first excitation gap vs $d$ data for several chain lengths. We start by noting two things: first, to go from one topological phase to another, a gapless phase (where spin-correlations decay algebraically with the system size) must be crossed. Therefore, one can use the algebraic decay of correlations to track $d_c$}; second, for $d = 0$ and $d\rightarrow-1$ \bluecolor{(the limit with dangling spin-1 at the edges)}, the gap to the first excited state decays exponentially with chain size \cite{kennedy1990exact} (see Fig. \ref{fig:decay_gap_0&crit_vs_N}a in Appendix \ref{Numerical determination of dc}). \bluecolor{The competition between the exponential and algebraic decays, produces a hump-like feature (see Fig. \ref{fig:dc_thermodynamic_limit}a in Appendix \ref{Numerical determination of dc})in the $d<0$ region of the gap vs $d$ plots, since the gap at $d=0$ and $d \rightarrow -1$ decreases exponentially with length, but only algebraically for $d = d_c$. Following the location of this feature as a function of chain length, and extrapolating to the thermodynamic limit, allows for the determination of the critical dimerization $d_c$. Even though we focused on the negative $d$ region, an identical phase transition occurs on the $d>0$ side for the same absolute value of dimerization. Extracting $d_c$ from the $d>0$ region is harder, since in that case the transition is between a gapless (Haldane) and gapped (dimerized without dangling spins) phase, without an easily identifiable feature to track. Additional details on this procedure are given in Appendix \ref{Numerical determination of dc}.} 

In what follows, we restrict our analysis to the region $\beta_1,\beta_2 < 1/3$ to avoid introducing additional phase transitions when compared with the case without biquadratic exchanges. This is also the relevant parameter region for nanographenes, since the biquadratic exchange has been shown to be one order of magnitude smaller than the bilinear interactions, i.e. $\beta \sim 0.1$ \cite{mishra2021observation,henriques2023anatomy}.

\begin{figure}
    \centering
    \includegraphics[width=\linewidth]{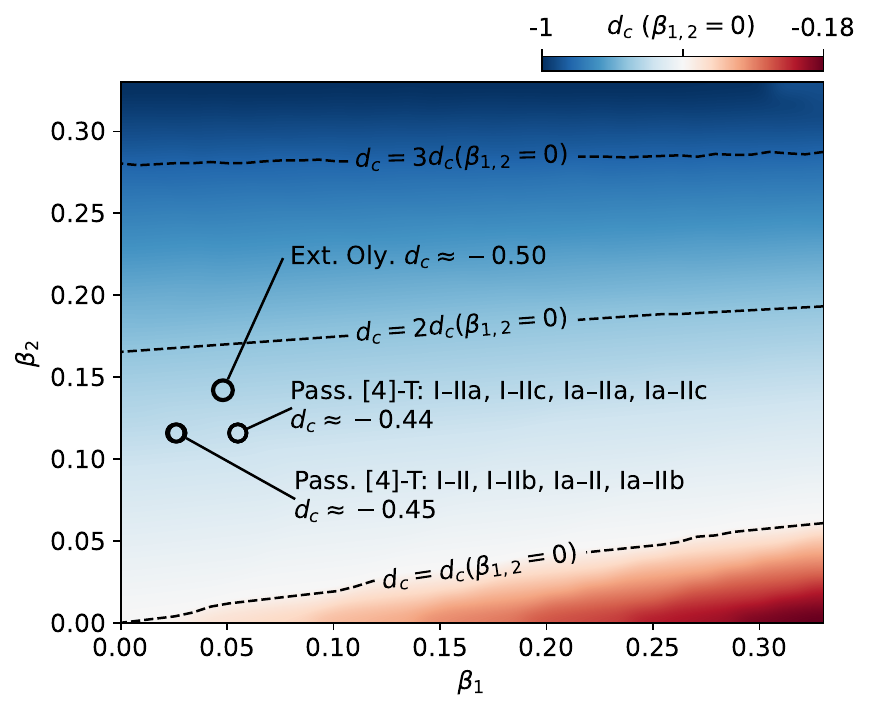}
    \caption{\textbf{Critical dimerization map.} Colormap of the critical dimerization value at which a topological phase transition occurs in the Hamiltonian of equation (\ref{Eq: S=1 BAH}) as a function of the relative strength of the biquadratic and bilinear exchanges, $\beta_{1,2} = B_{1,2} / J_{1,2}$ obtained with DMRG for chains with $N = 40$ $S=1$ spins. The reference value is  $|d_c(\beta_{1,2}=0)| \approx 0.277$.}
    \label{fig:dc_colormap}
\end{figure}

\bluecolor{Using the procedure outlined above, we solve the spin Hamiltonian using DMRG for chains with up to $N = 600$ spins, and find $|d_c| = 0.26$ for $\beta_1 = \beta_2 = 0$ (a value compatible with what was reported in Ref. \cite{kato1994numerical}).  In Fig. \ref{fig:40vs100_beta2}a of Appendix \ref{Numerical determination of dc} we show how calculations with $N = 40$ spins are sufficient to reproduce the dependence of $d_c$ observed in significantly larger systems, allowing for smaller computational costs. Repeating the procedure for $0 < \beta_{1,2} \leq 1/3$, and determining $d_c$ for each case, we find how the critical dimerization depends on the biquadratic exchanges. The result is depicted as a colormap in Fig. \ref{fig:dc_colormap}, with cross sections being given in Fig. \ref{fig:40vs100_beta2}b}

The DMRG results of Fig. \ref{fig:dc_colormap} show a marked dependence of the critical dimerization on \bluecolor{biquadratic} exchange. We find that in the considered $(\beta_1,\beta_2)$ region the critical dimerization parameters can range from $|d_c| \approx 0.18$ to $|d_c|\rightarrow 1$. Furthermore, we find that $d_c$ depends more strongly on $\beta_2$ (the one associated with the stronger exchange, since we considered the region $d<0$) than on $\beta_1$, and it approaches $|d_c| \rightarrow 1$ when $\beta_2 \rightarrow 1/3$, implying that for such value of $\beta_2$ the phase transition only takes place when the system becomes fully dimerized. On the other hand, for $\beta_1 \rightarrow 1/3$ and $\beta_2 \rightarrow 0$ a critical dimerization value \bluecolor{$|d_c| \approx 0.18$} is found. Along the line $\beta_1 = \beta_2 \equiv \beta$, and around the critical point $\beta_1 = \beta_2 = 1/3$, our results are consistent with the analytical result $d_c = (1+2\beta)/(3-4\beta)$ derived in \cite{totsuka1995isotropic}, as we show in Fig. \ref{fig:dc_num_vs_analytical} of Appendix \ref{Numerical determination of dc}. \bluecolor{To better understand some of these limit cases, we analyze the spin Hamiltonian expressed in terms of spin projectors in Appendix \ref{Analysis of dc}.}

\section{Nanographene building blocks}
\begin{figure*}
    \centering
    \includegraphics[width=\linewidth]{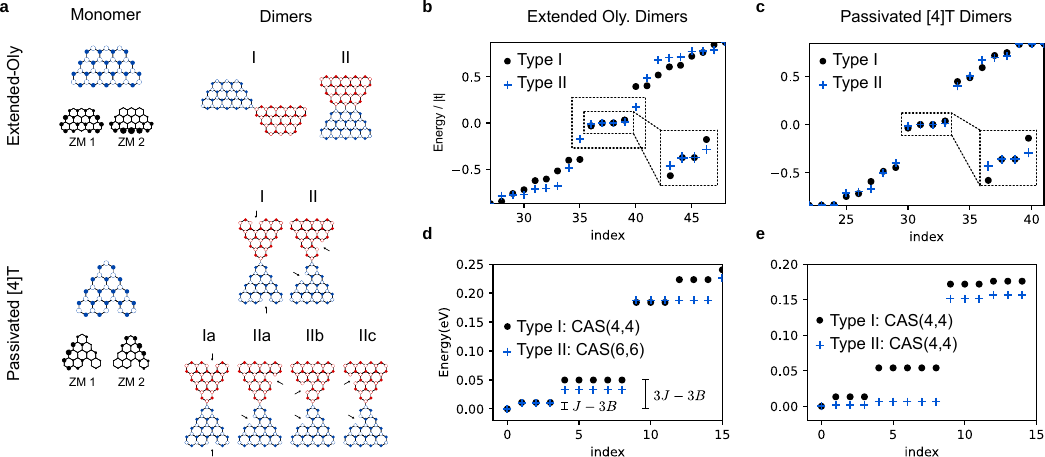}
    \caption{\textbf{Molecular building blocks.} \textbf{a}, $S=1$ nanographenes which can be used to realize chains described by the Hamiltonian of equation (\ref{Eq: S=1 BAH}), with the wave function of its zero modes (ZM) depicted below each molecule. Full (empty) circles mark the majority (minority) sublattice. Possible dimers obtained from these monomers are depicted on the right, where arrows indicate the passivated site. \textbf{b}, \textbf{c}, Single particle spectrum for the dimers of panel \textbf{a}. The dashed boxes highlight the active space used in the CI-CAS calculation for each system. \textbf{d}, \textbf{e}, Energies obtained by solving the Hubbard model in the CI-CAS approximation. The energy differences between the singlet and the triplet and quintuplet in the BLBQ model are also shown. The parameters $t = -2.7$eV, $t_3=t/10$ and $U = |t|$ were used.}
    \label{fig:monomer_dimers}
\end{figure*}

Having determined the dependence of $d_c$ with the biquadratic interactions, we move on to analyze  nanographenes which can be used to realize this Hamiltonian experimentally, and in which phase we expect the system to be: either in the Haldane phase of Ref. \cite{mishra2021observation}, or in the dimerized phase. In Fig. \ref{fig:monomer_dimers}a we depict two experimentally feasible molecules, each hosting two zero energy orbitals  and spin-1 ground state, that can be used to realize the desired model.

The extended-Oly. \cite{mishra2025synthesis} molecule is similar to the spin-1/2 Olympecene of Refs. \cite{mistry2015synthesis, zhao2024gapless}, where extra benzene rings are included, increasing the molecule's sublattice imbalance, and consequently its spin \cite{lieb1989two}. Like its $S=1/2$ counterpart, two of these molecules can be covalently linked to generate two types of dimers: one where the molecules are connected tip to tip (Type-I), and another where they are linked side to side (Type-II), forming an extended Clar's goblet \cite{clar1972circobiphenyl}. The latter has recently been synthesized on surface independently by two groups \cite{mishra2025synthesis, li2025designer}.

The other nanographene we consider is derived from the already synthesized [4]-triangulene \cite{mishra2019synthesis, su2020triangulenes} where a site along its edge is passivated 
through hydrogenation or dehydrogenation, a procedure already demonstrated experimentally \cite{li2019single, mishra2020topological, turco2023observation, zhao2024tailoring}. This passivation not only brings the ground state spin down from $S=3/2$ to $S=1$, as desired, but also breaks $C_3$ symmetry, creating an inhomogeneity in the spatial distribution of the molecular zero energy orbitals, which now accumulate away from the passivated site. This inhomogeneity is the key feature to produce the desired exchange alternation depending on how different units are covalently linked. We will be mainly focused on the Type-I and Type-II dimers of Fig. \ref{fig:monomer_dimers}a, and comment on the effect of using the Ia, IIa, IIb and IIc variations later. 

\section{Single particle and many-body spectra}
We describe the electronic states of these molecules with a Hubbard model, with first and third neighbor hoppings, $t=-2.7$eV, $t_3=0.1 t$  and on-site Hubbard interaction $U=|t|$ (see Appendix \ref{Extra CAS and TB} for details). We solve the many-body problem  with the Configuration Interaction in the Complete Active Space (CI-CAS) approximation, which we describe in a more detailed form in the Appendix \ref{Extra CAS and TB}. The validity of this approach and parameters has been successfully bench-marked in the past both with experiments and quantum chemistry calculations, including the case of nanographene-based topological spin chains \cite{ortiz2019exchange, mishra2021observation, catarina2023broken, zhao2024tunable}.  

\bluecolor{The first step to solve the Hubbard Hamiltonian in this approximation is to consider the tight-binding part of the Hamiltonian only (i.e. $U = 0$). In Fig. \ref{fig:monomer_dimers}b and c we show the tight-binding spectra for the Type-I and -II dimers of Fig. \ref{fig:monomer_dimers}a, with the results for the alternative dimers being shown in Fig. \ref{fig:Alternative dimers} of Appendix \ref{Extra CAS and TB}. In each single particle spectrum we identify four states close to zero energy, resulting from the hybridization of the two pairs of zero modes of the respective monomers, followed by an energy gap, and then the remaining states. The only exception to this is the extended-Oly. Type-II dimer where two additional orbitals can be identified close to the zero modes.}

\bluecolor{Importantly, inspection of the single-particle spectra already allows us to anticipate the modulation of the intermolecular exchange, since kinetic exchange scales with  the splitting of the zero modes \cite{jacob22,henriques2023anatomy}. Comparing the different spectra, we find that in the passivated [4]-triangulene dimers the difference in zero mode splitting is larger than in the extended-Oly. dimers. This is a consequence of the ingenious choice of passivated sites in the [4]-triangulene. By passivating sites close to the dimer's binding region, we effectively push the monomer's zero modes away from this vertex, leading to a smaller inter-triangulene mode hybridization \cite{jacob22, henriques2023anatomy}. The opposite is true when we passivate sites away from the binding region instead. As a result, we can expect a larger difference in the exchange parameters of the passivated [4]-triangulene system than in the extended-Oly. case.}

\bluecolor{Having solved the single particle problem, we consider now the case with finite $U$. To obtain the energy spectrum in this case we solve the Hamiltonian expressed in terms of a subset of single particle states; the active space. Here we adopt a minimal approach, where we solve the Hubbard model in a CAS(4,4) approximation containing only the zero modes for all dimers, except for the extended-Oly. Type-II dimer where we perform CAS(6,6) instead, in order to include the two orbitals which appear close to zero energy.} 

Depicted in Fig. \ref{fig:monomer_dimers}d and e is the result of the CAS(4,4) and CAS(6,6) calculations performed for the two sets of molecular dimers (results for the alternative dimers are shown in Fig. \ref{fig:Alternative dimers}). There, we find an energy spectrum which is split into a low and a high energy sectors. The former can be associated with spin-like states, while the latter has states where double occupancy of orbitals dominates. Matching the low energy part with the energy differences of the BLBQ model for a dimer \cite{catarina2023broken} (as depicted in Fig. \ref{fig:monomer_dimers}d), we obtain the bilinear, $J$, and biquadratic, $B$, exchanges displayed in Table I. As anticipated from the analysis of the single particle spectra, the exchange difference is larger between the Type-I and Type-II passivated [4]-triangulene dimers than in the ones obtained from extended-Oly. molecules.
\begin{table}[]
\begin{tabular}{|l|l|l|lll|}
\hline
\multicolumn{1}{|c|}{\multirow{2}{*}{Monomer}} & \multicolumn{1}{c|}{\multirow{2}{*}{Dimer}} & \multicolumn{1}{c|}{DFT}       & \multicolumn{3}{c|}{CAS}                                                  \\ \cline{3-6} 
\multicolumn{1}{|c|}{}                         & \multicolumn{1}{c|}{}                       & \multicolumn{1}{c|}{$J$ (meV)} & \multicolumn{1}{c|}{$J$ (meV)} & \multicolumn{1}{c|}{$B$ (meV)} & $\beta = B/J$ \\ \hline
\multirow{2}{*}{Extended-Oly}                  & I                                           & 14.01                             & \multicolumn{1}{l|}{19.36}     & \multicolumn{1}{l|}{2.74}      & 0.142   \\ \cline{2-6} 
                                               & II                                          & 12.15                             & \multicolumn{1}{l|}{11.69}     & \multicolumn{1}{l|}{0.56}      & 0.048\\ \hline
\multirow{6}{*}{Pass. {[}4{]}-T}                & I                                           & 16.49                           & \multicolumn{1}{l|}{20.36}     & \multicolumn{1}{l|}{2.36}      & 0.116   \\ \cline{2-6} 
                                               & II                                          & 2.51                            & \multicolumn{1}{l|}{2.27}      & \multicolumn{1}{l|}{0.06}      & 0.026   \\ \cline{2-6} 
                                               & Ia                                          & 16.51                          & \multicolumn{1}{l|}{20.37}     & \multicolumn{1}{l|}{2.36}      & 0.116   \\ \cline{2-6} 
                                               & IIa                                         & 5.03                            & \multicolumn{1}{l|}{5.45}      & \multicolumn{1}{l|}{0.3}       & 0.055   \\ \cline{2-6} 
                                               & IIb                                         & 2.87                             & \multicolumn{1}{l|}{2.33}      & \multicolumn{1}{l|}{0.06}      & 0.026   \\ \cline{2-6} 
                                               & IIc                                         & 5.05                             & \multicolumn{1}{l|}{5.46}      & \multicolumn{1}{l|}{0.3}       & 0.055   \\ \hline
\end{tabular}

\label{tab: exchange}
\caption{Summary of the bilinear ($J$) and biquadratic ($B$) exchanges obtained by matching the energies of the BLBQ Hamiltonian for the dimers of Fig. \ref{fig:monomer_dimers} and the results obtained from CI-CAS; the strength of \bluecolor{biquadratic} exchange is quantified with $\beta = B/J$. The DFT exchanges were obtained from the energy difference between the ferromagnetic and antiferromagnetic configurations.}
\end{table}

\bluecolor{In Appendix. \ref{Extra CAS and TB} we study how the CI-CAS results change as more orbitals are included in the active space, and how the results depend on $U$. We find that the results presented in Table I do not change drastically when more orbitals are included or slightly different values of $U$ are used. Nonetheless, to strengthen the validity of our results, we use DFT (details of the calculation are given in Appendix \ref{Extra DFT Calc}) \cite{QE-2017, QE-2009, doi:10.1063/5.0005082, PhysRevLett.43.1494, PhysRevB.43.1993, van_Setten_2018, Hamann2013, garrity2014pseudopotentials, doi:10.1126/science.aad3000} to compute the \bluecolor{bilinear} exchange from the energy difference between the ferromagnetic and antiferromagnetic solutions for each dimer. The results are also shown in Table I, and are in good agreement with the Hubbard model calculation. In the same appendix, we extend our analysis to trimer, and explain why second neighbor exchange or three spin terms can be safely ignored from our model Hamiltonian in these systems.}

\begin{table}[]
\begin{tabular}{clllll}
\cline{2-3}
\multicolumn{1}{c|}{\multirow{2}{*}{Extended-Oly}}    & \multicolumn{1}{l|}{$|d|$} & \multicolumn{1}{l|}{II}   &                           &                           &                           \\ \cline{2-3}
\multicolumn{1}{c|}{}                                 & \multicolumn{1}{l|}{I}   & \multicolumn{1}{l|}{0.25} &                           &                           &                           \\ \cline{2-3}
                                                      &                          &                           &                           &                           &                           \\ \cline{2-6} 
\multicolumn{1}{c|}{\multirow{3}{*}{Pass. {[}4{]}-T}} & \multicolumn{1}{l|}{$|d|$} & \multicolumn{1}{l|}{II}   & \multicolumn{1}{l|}{IIa}  & \multicolumn{1}{l|}{IIb}  & \multicolumn{1}{l|}{IIc}  \\ \cline{2-6} 
\multicolumn{1}{c|}{}                                 & \multicolumn{1}{l|}{I}   & \multicolumn{1}{l|}{0.8}  & \multicolumn{1}{l|}{0.58} & \multicolumn{1}{l|}{0.79} & \multicolumn{1}{l|}{0.58} \\ \cline{2-6} 
\multicolumn{1}{c|}{}                                 & \multicolumn{1}{l|}{Ia}  & \multicolumn{1}{l|}{0.8}  & \multicolumn{1}{l|}{0.58} & \multicolumn{1}{l|}{0.8}  & \multicolumn{1}{l|}{0.58} \\ \cline{2-6} 
\end{tabular}
\label{tab: dimerization}
\caption{Summary of the dimerizations $d$ computed using the \bluecolor{bilinear} exchanges given in Table I for the extended-Oly. and the passivated [4]-triangulene dimers. These values should be compared with the critical dimerization values displayed in Fig. \ref{fig:dc_colormap}.}
\end{table}

In Table II we summarize the dimerizations associated with every possible pair of \bluecolor{bilinear} exchanges obtained from CI-CAS.  We find dimerization values between $ |d| = 0.58$ and $|d|=0.8$ for the passivated [4]-triangulene (depending on the kind of Type-I and Type-II dimers), and $|d|=0.25$ for the extended-Oly. case. By comparing  the dimerizations of Table II with the critical dimerization values highlighted in the colormap of Fig. \ref{fig:dc_colormap}, we can place the chains made from these building blocks in the Haldane or dimerized phases. Doing this, we find that a chain made from the passivated [4]-triangulene should be in the dimerized phase given that, for the relevant $(\beta_1,\beta_2)$, the dimerization is always larger than the critical values $|d_c|=0.44$ and $|d_c| = 0.45$. Such a chain will have a ground state which is either non-degenerate or ninefold degenerate in the thermodynamic limit, depending on whether the chain ends on a weak or strong bond. In contrast, chains built from extended-Oly dimers should fall in the Haldane phase due to their small dimerization, which is well below the critical value $|d_c| = 0.5$. For long enough chains a fourfold degenerate ground state is expected, similarly to what was found with the [3]-triangulene chains \cite{mishra2021observation}. \bluecolor{Notice how computing the $d_c$ dependence with $(\beta_1,\beta_2)$ was crucial for the extended-Oly. case, where the dimerization $|d| = 0.25$ would place the system right at the phase transition if one considered the $|d_c ( \beta_1 = 0, \beta_2=0)| = 0.25$ obtained in \cite{kato1994numerical} as the reference value.}

\section{IETS simulation}

As it was shown in previous works, one way to directly observe topological phases in spin chains is through inelastic electron tunneling spectroscopy (IETS) \cite{mishra2021observation, zhao2024tunable, zhao2024gapless, yuan2025fractional,su2024fabrication,sun2025surface}. Due to the atomic resolution of STM, this techniques allows the measurement of spin excitations throughout the chain, making it possible to identify bulk properties (in the middle of the chain) as well as the presence of topological states (at the edges). Here we simulate the IETS spectrum using perturbation theory \cite{Rossier2009ITS} up to third order in the tip-surface tunneling \cite{ternes2015spin} (see Appendix \ref{dIdV details}), for spin chains described with the Hamiltonian of equation (\ref{Eq: S=1 BAH}) using the parameters of Table I.
\begin{figure*}
    \centering
    \includegraphics[width = \linewidth]{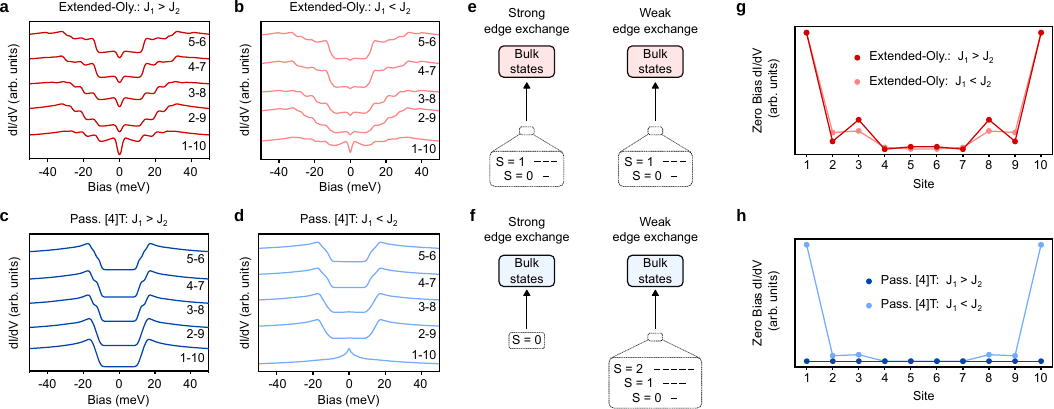}
    \caption{\textbf{Simulation of IETS.} \textbf{a}, \textbf{b}, \textbf{c}, \textbf{d}, Simulation of the $dI/dV$ maps for chains with 10 sites described with equation (\ref{Eq: S=1 BAH}), made from extended-Oly. and passivated [4]-triangulenes, using the exchanges of Table I. \textbf{e}, \textbf{f}, Pictorial representation of the spectrum of the spin Hamiltonian for the two types of chains with different terminations, i.e. chains starting with a strong or weak exchange. \textbf{g}, \textbf{h}, $dI/dV$ signal at zero bias across the chain obtained from panels \textbf{a}, \textbf{b}, \textbf{c}, \textbf{d}.}
    \label{fig:Spin Chains IETS}
\end{figure*}

We start by modeling the extended-Oly. chains. To do so, we first build the Hamiltonian using the exchange values of Table I, we diagonalize it exactly \cite{weinberg_quspin_2017} and finally compute the conductance ($dI/dV$) as a function of the applied bias for each site of a chain with 10 sites. In Fig. \ref{fig:Spin Chains IETS}a and b, we show the simulated $dI/dV$ for the case where the first exchange in the chain is the largest one, $J_1 > J_2$ (the first dimer in the chain is Type-I), and for $J_1 < J_2$ (the first dimer is Type-II). In both cases we find similar results, with the most important feature being the singlet to triplet excitation near zero bias (see Fig. \ref{fig:Spin Chains IETS}e), whose spectral weight decays exponentially into the chain. In the thermodynamic limit, the singlet and triplet become degenerate, and this feature evolves into a Kondo peak. This finite signal at zero bias is the fingerprint of the Haldane phase, as discussed previously in \cite{mishra2021observation}. We note that for a chain with $N=10$ spins, the singlet-triplet splitting is still finite, and the Kondo peak is not yet fully formed, in line with the experiments in [3]-triangulenes \cite{mishra2020topological} with chains with less than 8 units. Nonetheless, we can still clearly see, for both cases, the edge-nature of this excitation by tracking the \bluecolor{value of the conductance at zero bias along all sites}, as we depict in Fig. \ref{fig:Spin Chains IETS}g.   

In Fig. \ref{fig:Spin Chains IETS}c and d we show the simulation for the chains made with passivated [4]-triangulenes (using the exchange values of the Type-I and -II dimers of Table I). For $J_1 > J_2$ the system is expected to be in the fully dimerized phase with a singlet ($S=0$) ground state. This is clearly seen in the IETS spectrum from the presence of an excitation gap across all sites. If instead we consider the case where the first exchange is the weak one,  $J_1 < J_2$, then we are at the opposite side of the phase diagram given in Fig. \ref{fig:Spin Chains}d. One sees this in the $dI/dV$ simulation due to the presence of Kondo peaks (this time fully developed) at the edge units, resulting from the emergent spin-1 at the chain edges, leading to nearly degenerate $S=0,1,2$ manifolds (see Fig. \ref{fig:Spin Chains IETS}f). Tracking the height of the zero bias conductance throughout the chain (as shown in Fig. \ref{fig:Spin Chains IETS}h), a stark difference is seen between these two phases: when the first exchange is the strongest one, no signal is obtained at zero bias, consistent with a gapped system; alternatively, when the first exchange is the weakest one, then the signal peaks at the edges and decays exponentially into the chain, with slightly stronger signal at the odd sites. 

Hence, we see that IETS provides a clear way to detect whether the chains are in the Haldane or the dimerized phase. For the former, Kondo peaks (or low energy spin excitations) are expected at the edges, regardless of the type of exchange at the chain edges. For the latter, only the chains terminated in weak exchange will show low energy features.

\section{Robustness of dimerized phase}
When trying to build chains whose building blocks are the passivated [4]-triangulenes, it might not be possible to control the exact location for passivation due to the equivalence of the two middle sites along an edge. Thus, while the edge along which passivation is performed can in principle be selected, the specific site that is going to be passivated is harder to control. As a result, we expect that the type of dimers might change throughout the chain, leading to bond dependent values of $d$, $\beta_1$ and $\beta_2$. As can be seen from Table I, both Type-I and Type-Ia dimers have almost identical exchanges both in CI-CAS and DFT. Likewise, we find that Type-II and Type-IIb, as well as Type-IIa and Type-IIc, are pairwise very close to each other, although slight differences are found between the pairs. 

To check if these random exchange variations can affect the topological phase of our system, we once again use DMRG \cite{fishman_itensor_2022} to solve the Hamiltonian for a chain with $N=40$ spins, terminated on a strong bond (which has a non-degenerate ground state in the thermodynamic limit), and check the size of the gap between the ground state and the first excited state as we increase the disorder in the system. The degree of disordered is controlled by a parameter $p$. \bluecolor{For a given strong bond, we consider Type-I and Type-Ia exchanges with probabilities $1-p/2$ and $p/2$, respectively; for the weak bonds we consider Type-II exchange with probability $1-3p/4$, and Type-IIa, Type-IIb and Type-IIc with probability $p/4$ each.} This way, when $p = 0$, the exchanges alternate between Type-I and Type-II along the chain, and when $p=1$ the Type-I and Type-Ia exchanges can occur with equal probability, and the same for the Type-II, IIa, IIb and IIc values. The latter scenario is the one which might more closely describe what is expected to happen experimentally. 

\begin{figure}
    \centering
    \includegraphics[width=\linewidth]{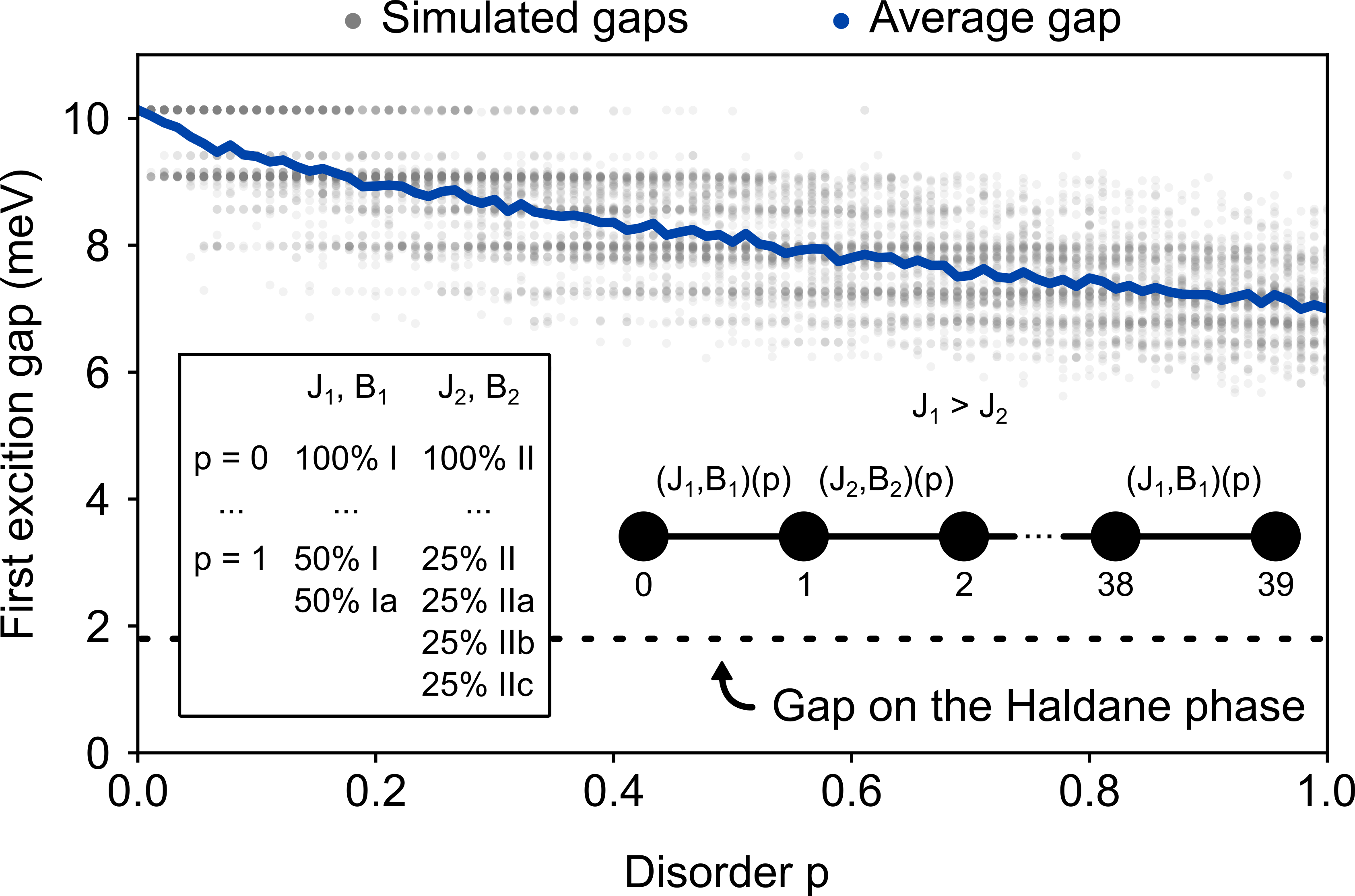}
    \caption{\textbf{Disorder dependence.} Gap to the first excited state for a chain starting on a strong exchange as a function of the disorder parameter $p$ defined in the text. For each $p$ a chain with $N = 40$ spins is diagonalized $100$ times, with each gap being represented by a semi-transparent black circle. The blue line represents the average gap for each $p$. The dashed line shows the gap at the onset of the Haldane phase without disorder (changing $J_2$ such that $d = d_c = 0.44$) which is only finite due to finite-size effects.} 
    \label{fig: gap vs p}
\end{figure}

In Fig. \ref{fig: gap vs p}, for each value of $p$ we diagonalize the system 100 times and track the magnitude of the gap.
There, we see that, as disorder increases, the mean gap value steadily decreases, but never vanishes. For a more accurate analysis, in Fig. \ref{fig: gap vs p}, we also show a dashed line marking the gap in the case where $J_2$ is modified such that the dimerization in the chain is $|d| = |d_c| = 0.44$, and no disorder is considered. i.e. $p = 0$. For an infinite chain this gap would vanish (since that is the onset of the Haldane phase), and the finite value we find is only due to finite size effects. Crucially, one sees that the average gap for the original spin chain always remains significantly larger than this reference gap, showing that the disorder introduced by the variation of dimers across the chain does not take the system away from the topological dimerized phase. This is an important result for the future realization of this system, as it shows that the passivated [4]-triangulene chains are resilient to perturbations in the exchange couplings and present themselves as an excellent candidate to realize the $S = 1$ analog of the even- and odd-Haldane phases found with $S=1/2$ Olympicenes experimentally \cite{zhao2024tunable}.

\section{Conclusion}
\bluecolor{In summary, we are proposing an experimental route to probe both the Haldane and dimerized phases predicted to exist in bond alternating spin-1 chains} \cite{totsuka1995isotropic,kitazawa1996phase}. We propose to use $S=1$ nanographenes as building blocks for the spin chain, a route that has already shown its versatility for the simulation of  other model Hamiltonians \cite{mishra2021observation,zhao2024gapless,zhao2024tunable}. 
Given that \bluecolor{biquadratic} exchange is not negligible in nanographenes, we extend previous theory work to analyze the quantum phase transition of the $S=1$ BAH Hamiltonian to the case with alternating \bluecolor{biquadratic} exchange. We then have proposed two specific families of molecules realizing both sides of the quantum phase transition and we made specific predictions of how the IETS could be used to probe them.

Our proposed route to explore this Hamiltonian is superior to prior works with bulk samples and a macroscopic number of chains \cite{hagiwara1998experimental, narumi2001evidence, narumi2004high, zheludev2004distribution, hagiwara2005spin, hagiwara2006tomonaga,madhumathy2023crystal} on several counts: 
First, the use of nanographenes as building blocks for artificial spin chains allows for a direct measurement of the exchange interactions in dimers using IETS and comparing with theory. 
Second, a direct spectral image of the edge excitations can be probed with this method.
Third, using tip-induced (de-) hydrogenation it is possible to selectively passivate/activate spin units in a given spin chain \cite{zhao2024gapless}, allowing for an unprecedented degree of control over the system's properties.
Fourth, intermolecular exchange in nanographenes is large, in the order of tens of meV, and several orders of magnitude superior to magnetic anisotropy \cite{lado2014magnetic}. 
Therefore, nanographenes made through on-surface synthesis provide a great platform for the exploration of quantum magnetism in artificial spin lattices. Our work proposes a ready to test set of experimental predictions that can be used to explore one-dimensional quantum magnetism.

\section*{Acknowledgements}
The authors thank Gonçalo Catarina for fruitful discussions.

\paragraph{Author contributions}
J.H. and J.F.R. supervised the project and coordinated the collaborations. 
Y.C. carried out the DMRG calculations. 
J.H. and R.S. solved the tight-binding and Hubbard models. 
J.P. was responsible for the DFT calculations.
J.H. performed the IETS simulation.
J.H. and J.F.R. wrote the manuscript with contributions from the other authors.

\paragraph{Funding information}
J.H. and J.P. acknowledge financial support from
SNF Sinergia (Grant Pimag).
J.F.R.  acknowledges financial support from 
FCT (Grant No. PTDC/FIS-MAC/2045/2021),
SNF Sinergia (Grant Pimag),
Generalitat Valenciana funding Prometeo2021/017
and MFA/2022/045, and funding from MICIIN-Spain (Grant No.  PID2022-141712NB-C22).
Y.C. acknowledges funding from 
FCT, QPI,  (Grant No.SFRH/BD/151311/2021) and 
thanks the hospitality of the Departamento de F\'isica Aplicada at the Universidad de Alicante.


\begin{appendix}
\numberwithin{equation}{section}


\section{Numerical determination of $d_c$}
\label{Numerical determination of dc}
Here we discuss the numerical details to obtain the critical dimerization $d_c$ for different $\beta_1$, $\beta_2$ by tracking the energy gap between the ground state and the first excited state as a function of the dimerization parameter $d$ for open chains of length $N$.
 
\begin{figure}[h]
    \centering
    \includegraphics[width=0.65\linewidth]{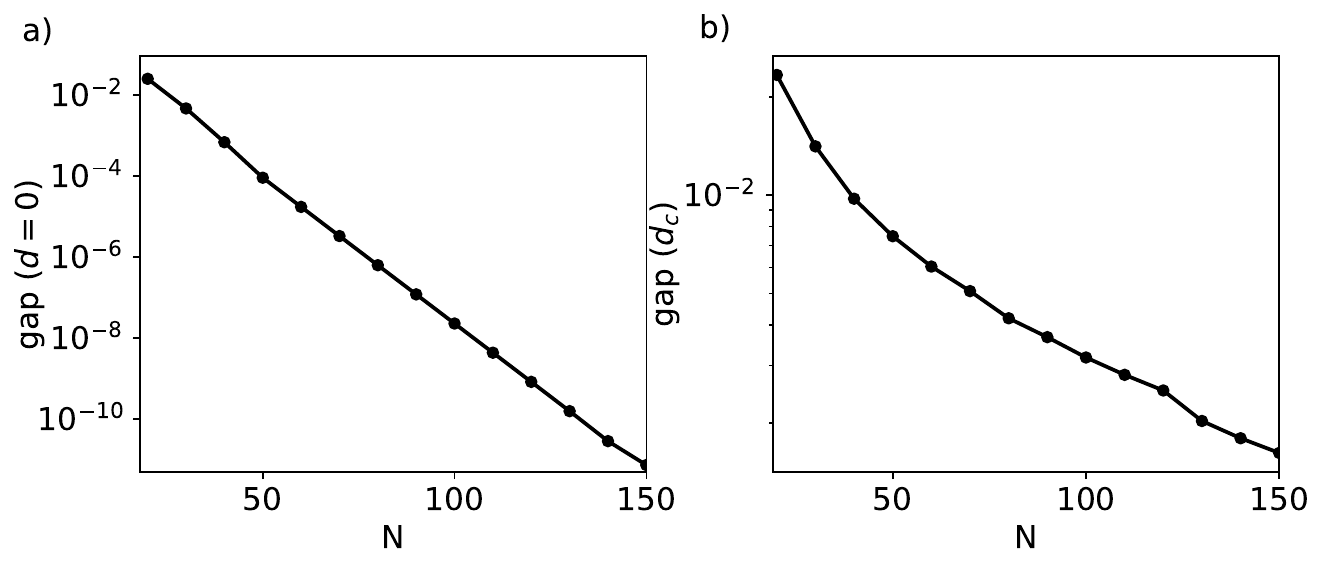}
    \caption{Panel a) and b) show the decay of the energy gap between the ground state and the first excited state as a function of the chain length $N$ for $d=0$ and $d=d_c$, respectively, with $\beta_{1,2}=0$. \bluecolor{Energy gaps are normalized by the exchange coupling $J = J_1 = J_2$.}}
    \label{fig:decay_gap_0&crit_vs_N}
\end{figure}

For open boundary conditions, the spectrum is not symmetric under $d \!\to\! -d$. \bluecolor{For $d \geq 0$, we start in the Haldane phase ($d=0$) where the ground state is a singlet followed by a triplet whose energy decays exponentially with chain length, leading to a fourfold degeneracy in the thermodynamic limit (Fig. \ref{fig:decay_gap_0&crit_vs_N}a). When $d = d_c$ the system becomes gapless in the thermodynamic limit but with an algebraic decay of the first excitation gap (Fig. \ref{fig:decay_gap_0&crit_vs_N}b). For $d > d_c$ we enter the dimerized phase, the ground state is a singlet, and the first excited state is a triplet whose energy approaches a finite value as the chain length increases. For $d < 0$, we start once again in the Haldane phase, and progress to the gapless phase at $d = d_c$. When $d < d_c$ we enter a dimerized phase with dangling spins, where the first excitation gap decays exponentially with system size, producing a ninefold degenerate ground state in the thermodynamic limit.}

\begin{figure*}[h!]
    \centering    \includegraphics[width=0.9\columnwidth]{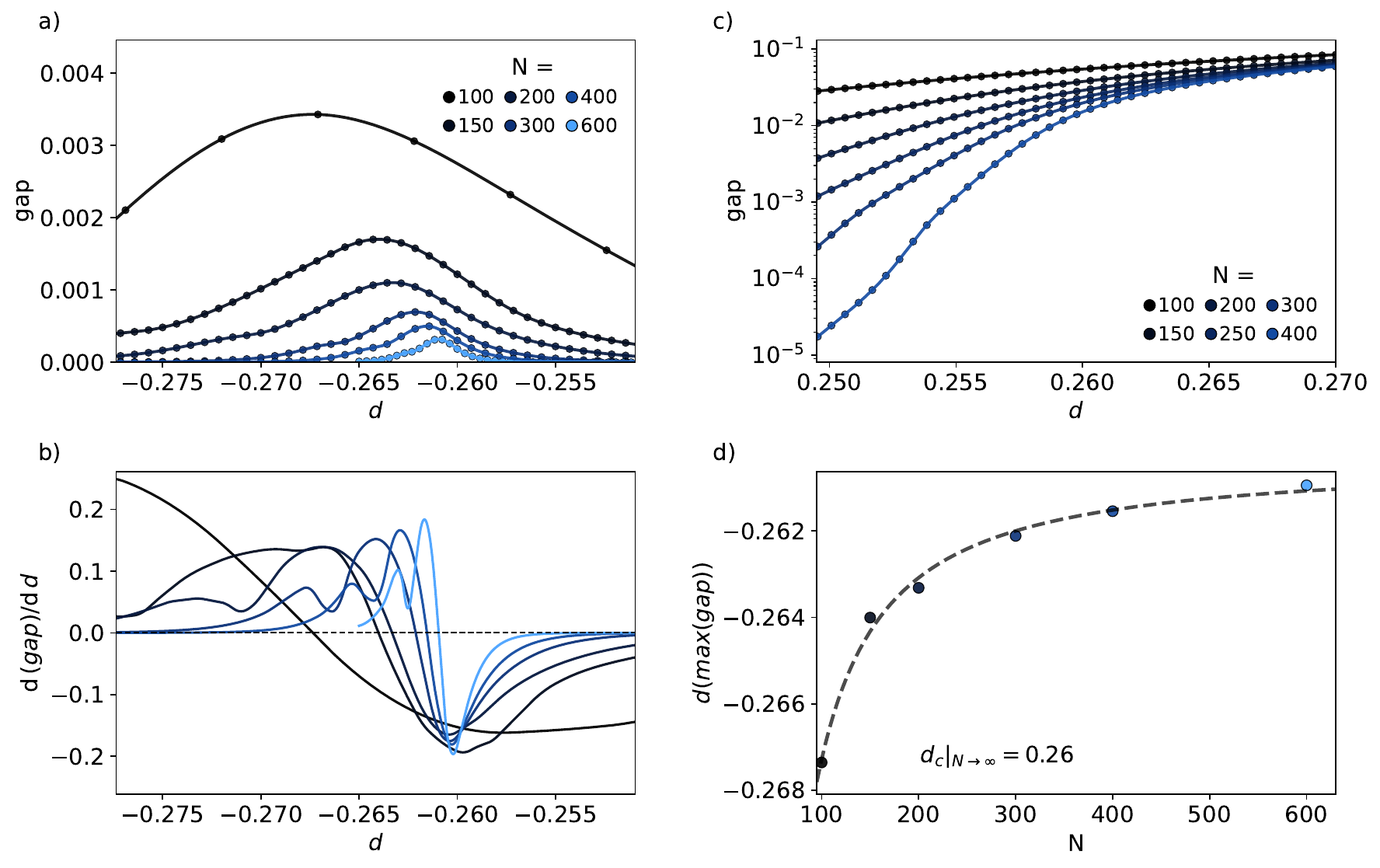}
    \caption{Panels a) and c) show the energy gap between the ground state and the first excited state as a function of the dimerization parameter $d$ for negative and positive values respectively, and for different chain lengths $N$. In panel a), the small hump indicates the critical dimerization value $d_c$. Panel b) shows the derivative of the data in panel a), which makes the convergence of the critical point more evident. Panel d) shows the position of the hump \bluecolor{in panel a)}, i.e. the value of $d$ at which the gap is maximal, as a function of $N$. The data are fitted to a polynomial scaling law, yielding a critical value in the thermodynamic limit of $d_c = 0.26$. \bluecolor{In all panels $\beta_1 = \beta_2 = 0$, $d = (J_1 - J_2)/(J_1 + J_2)$ and energy gaps are given in units of $J = (J_1 + J_2)/2$.}
 }
    \label{fig:dc_thermodynamic_limit}
\end{figure*}

\bluecolor{To obtain the value of $d_c$ we take advantage of the different decay rates at which the first excitation gap evolves with chain length. For negative $d$, we expect a first excitation gap which vanishes in the entire $-1 < d < 0$ region as the chain length increases. Crucially, this gap will approach zero at a slower rate at the critical dimerization value $d_c$ due to the algebraic correlations of the gapless phase, leading to a hump-like feature in the first excitation gap vs $d$ plots (Fig. \ref{fig:dc_thermodynamic_limit}a and b). Even though a critical dimerization exists for both $d>0$ and $d<0$, we find no easily identifiable feature for $d>0$ (Fig. \ref{fig:dc_thermodynamic_limit}c). The hump at negative $d$ is easy to track, and yields $d_c$ when extrapolated to the thermodynamic limit (Fig. \ref{fig:dc_thermodynamic_limit}d).}

All DMRG calculations were performed with \texttt{ITensors.jl} \cite{fishman_itensor_2022}. For each parameter set $(\beta_1,\beta_2,d)$ and chain length $N$, we targeted the lowest-energy states in the total-spin sectors $S=0$ and $S=1$. The number of sweeps  and the maximum bond dimension $m$ were scaled with $N$, and in the largest system, $N=600$, corresponds to about 40 sweeps with $m_{\max}\approx 420$. Within each run, the bond dimension was gradually increased from $m=50$ to $m_{\max}$, while the truncation cutoff was reduced from $10^{-9}$ to $10^{-12}$ across the sweeps.

Panels a) and c) of Fig. \ref{fig:dc_thermodynamic_limit} show the gap for $d<0$ and $d>0$, respectively, for several chain lengths up to $N=600$.
Panel b) shows the derivative of the negative-$d$ data, marking the evolution of the maximum in the hump as $N$ increases. For each $N$ we extract the position $d^*(N)$ of the hump (the value of $d$ at which the gap is maximal) and extrapolate it to the thermodynamic limit using $d_c(N) \;=\; d_c|_{N\rightarrow \infty} \;+\; \frac{a}{N}+ \frac{b}{N^2},$
which yields $d_c|_{N\rightarrow \infty} = 0.26$, as shown in panel d) of Fig. \ref{fig:dc_thermodynamic_limit}.

\begin{figure}[h]
    \centering
    \includegraphics[width=0.9\linewidth]{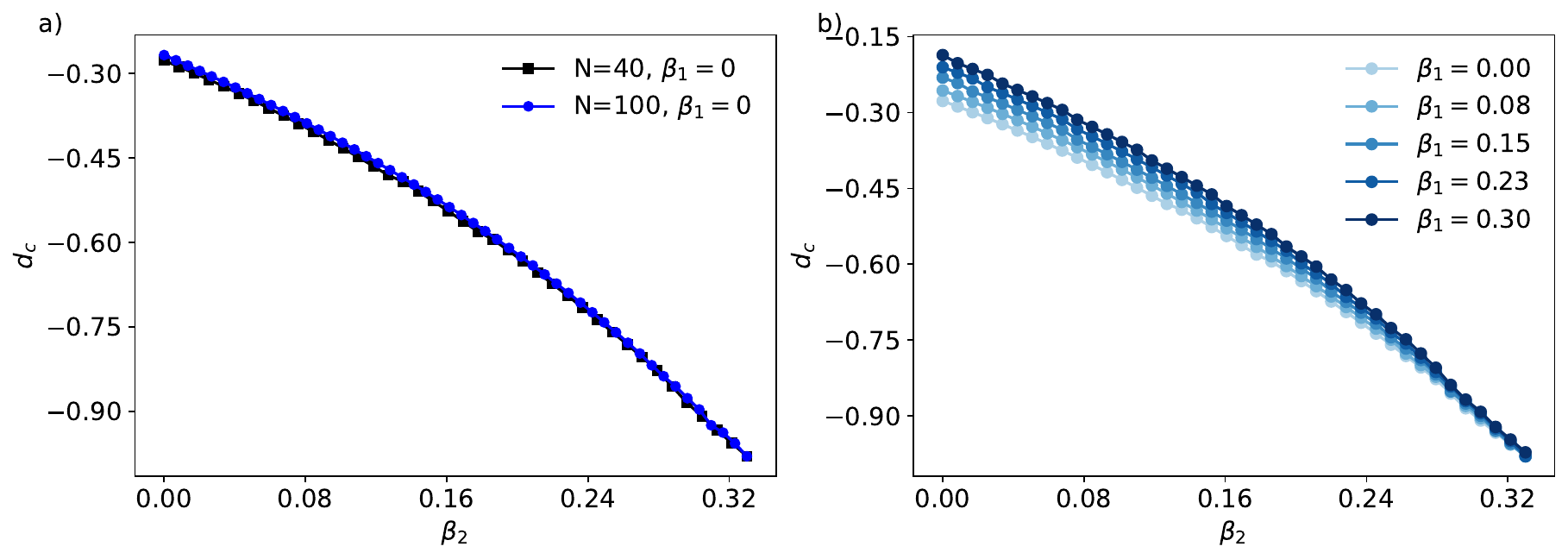}
    \caption{a) Comparison of the evolution of $d_c$ as a function of $\beta_2$ for a fixed $\beta_1=0$ between two chain lengths of $N=40$ and $N=100$. b) Variation of $d_c$ with $\beta_2$ at different fixed $\beta_1$ values for a chain of size $N=40$.}
    \label{fig:40vs100_beta2}
\end{figure}

In the colormap representation of the phase diagram \bluecolor{(Fig. \ref{fig:dc_colormap})}, we show the critical dimerization value $d_c$ as a function of $\beta_1$ and $\beta_2$. For this calculation we used a chain of length $N=40$, for which we find $d_c (\beta_{1,2}=0) \approx 0.277$. The choice of $N=40 $ is motivated by computational efficiency, since larger systems are more costly to simulate with DMRG. To verify that this choice is sufficient, we compared the evolution of $d_c$ as a function of $\beta_2$ at a fixed $\beta_1=0$ for system sizes, $N=40$ and $N=100$. As shown in Fig. \ref{fig:40vs100_beta2}, the results are close enough to capture the same evolution of $d_c$. In addition, Fig. \ref{fig:40vs100_beta2}b shows cross-sections of the colormap in Fig. \ref{fig:dc_colormap} for several fixed values of $\beta_1$. These cross-sections make it clear that the strong-link $\beta$, in this case $\beta_2$ (since $d<0$) governs the overall evolution of $d_c$, while the weak-link parameter $\beta_1$, produces a slight shift of the hump position. 

The colormap in Fig. \ref{fig:dc_colormap} was obtained from a $40\times40$ grid in $(\beta_1,\beta_2)$  using chains of length $N=40$. For each point in parameter space the gap was computed along a moving window of 20 values of $d$ centered around the expected position of the hump, with DMRG runs performed using 20 sweeps and a maximum bond dimension of $m_{\max}=80$, while the truncation cutoff was gradually reduced from $10^{-8}$ to $10^{-12}$ across the sweeps.

\begin{figure}
    \centering
    \includegraphics[width=0.5\linewidth]{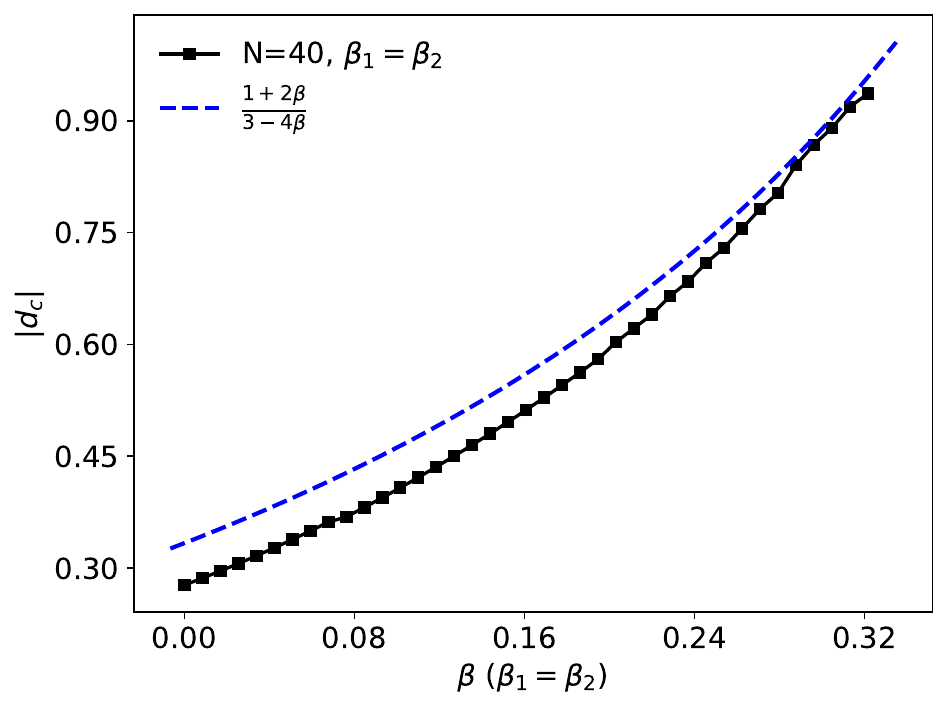}
    \caption{Comparison between numerical and approximate analytical expression for $d_c$ as a function of $\beta_1 = \beta_2 \equiv \beta$.}
    \label{fig:dc_num_vs_analytical}
\end{figure}

In Fig. \ref{fig:dc_num_vs_analytical} we show the comparison between the $d_c$ found numerically, and the approximate analytical expression found in \cite{totsuka1995isotropic} for the specific case with $\beta_1 = \beta_2$; we find a good agreement between the two approaches.
%


\section{Analysis of critical dimerization $d_c$ in special cases}
\label{Analysis of dc}

In this section we will look in more detail at some special cases of the BAH $S=1$ Hamiltonian in order to better understand the colormap of the critical dimerization value as a function of $\beta_1$ and $\beta_2$ shown in Fig. \ref{fig:dc_colormap}. We start by writing again the Hamiltonian presented in the main text:
\begin{align}
    H^{S=1}_\textrm{BAH} & = \sum_{i=0} J_1 \boldsymbol{S}_{2i} \cdot \boldsymbol{S}_{2i+1} + J_2 \boldsymbol{S}_{2i+1} \cdot \boldsymbol{S}_{2i+2}
    + \sum_{i=0} B_1 (\boldsymbol{S}_{2i} \cdot \boldsymbol{S}_{2i+1})^2 + B_2(\boldsymbol{S}_{2i+1} \cdot \boldsymbol{S}_{2i+2})^2 \label{Eq: S=1 BAH app}.
\end{align}
If one defined the dimerization as $d = (J_1 - J_2)/(J_1 + J_2)$, $J = (J_1 + J_2)/ 2$ and $\beta_{1,2} = B_{1,2}/ J_{1,2}$ this can be written as
\begin{align}
    \frac{H^{S=1}_\textrm{BAH}}{J} & = \frac{1 + d}{2} \sum_{i=0} \boldsymbol{S}_{2i} \cdot \boldsymbol{S}_{2i+1} + \beta_1 (\boldsymbol{S}_{2i} \cdot \boldsymbol{S}_{2i+1})^2 
    + \frac{1 - d}{2} \sum_{i=0} \boldsymbol{S}_{2i + 1} \cdot \boldsymbol{S}_{2i+2} + \beta_2 (\boldsymbol{S}_{2i + 1} \cdot \boldsymbol{S}_{2i+2})^2 .
\end{align}

First, we recall that in \cite{kato1994numerical}, where the system with $\beta_1 = \beta_2 =0$ and finite $d$ was considered, it was found that the degeneracy of the ground state would change (and thus a phase transition would occur) for a critical value of dimerization $d_c = 0.25 \pm 0.01$. Next, we note that for $d = 0$, the system's ground state in the thermodynamic limit was numerically verified to always be 4-fold degenerate for $0 \leq \beta_{1,2} \leq 1/3$. This is not surprising, as Kennedy \cite{kennedy1990exact} shows this to be the case when $\beta_1 = \beta_2$, and it will be used as the reference ground state in the following analysis. In the following, we will look at 3 other particular choices of $\beta_1$ and $\beta_2$, and for each we will study its ground state in the $d = -1$ limit. If the ground state is the same as in the $d = 0$ case we conclude that no phase transition happens in the system as we go from an isotropic exchange to a fully dimerized picture. On the other hand, if two different ground states are found, then a phase transition occurred at some critical dimerization value $|d_c| \in [0,1)$.

In the limit where $\beta_1 = \beta_2 = 1/3$, the Hamiltonian can be written in terms of projectors onto states with spin 2 \cite{affleck1988valence}, and apart from same constant energy shift it reads
\begin{align}
    \frac{H_\textrm{BAH} ^ {S=1}}{J} &= \frac{1 + d}{2} \sum_{i=0} 2 P_{S = 2} (2i, 2i + 1)
    + \frac{1 - d}{2} \sum_{i=0} 2 P_{S = 2} (2i + 1, 2i + 2)
\end{align}
where $P_{S = 2} (i,j)$ is the projector operator onto spin 2 states acting on sites $i$ and $j$. For $d = -1$ the Hamiltonian simply reads:
\begin{align}
    \lim_{d \rightarrow -1} \frac{H_\textrm{BAH} ^ {S=1}}{J} = \sum_{i=0} 2 P_{S = 2} (2i + 1, 2i + 2).
\end{align}
Since projectors are positively defined, a state with zero energy is necessarily a ground state of this Hamiltonian. The AKLT wave function \cite{affleck1988valence}, by construction, is an eigenstate of this Hamiltonian with zero energy, thus making it the ground state. Therefore, when going from $d = 0$ and $d = -1$ the ground state of the system remains unchanged, and no phase transition is expected. This is consistent with the colormap of Fig. \ref{fig:dc_colormap}, where it was found that when $\beta_{1,2} \rightarrow 1/3$, $|d_c| \rightarrow 1$, since all the phase space is take by the 4-fold degenerate Haldane phase, and the critical dimerization value is pushed to its maximum value.

For $\beta_1 = 0$ and $\beta_2 = 1/3$, the Hamiltonian (up to a constant energy shift) reads
\begin{align}
    \frac{H_\textrm{BAH} ^ {S=1}}{J} &= \frac{1 + d}{2} \sum_{i=0} \boldsymbol{S}_{2i} \cdot \boldsymbol{S}_{2i + 1} 
    + \frac{1 - d}{2} \sum_{i=0} 2 P_{S = 2} (2i + 1, 2i + 2).
\end{align}
In the limit $d \rightarrow -1$ we once again recover the Hamiltonian of the previous paragraph, and therefore the some conclusions apply. Therefore, we find that along the line $\beta_2 = 1/3$ no phase transition happens, in agreement with the result depicted in the colormap of Fig. \ref{fig:dc_colormap}, where in that region the critical dimerization value is always found to approach $|d_c| \rightarrow 1$.

At last, we look at the case with $\beta_2 = 0$ and $\beta_1 = 1/3$. The Hamiltonian reads:
\begin{align}
    \frac{H_\textrm{BAH} ^ {S=1}}{J} &= \frac{1 + d}{2} \sum_{i=0} 2 P_{S = 2} (2i, 2i + 1) 
    &+ \frac{1 - d}{2} \sum_{i=0} \boldsymbol{S}_{2i + 1} \cdot \boldsymbol{S}_{2i + 2}.
\end{align}
For $d = -1$, the term with the spin projector vanishes, and one is left with 
\begin{align}
    \lim_{d \rightarrow -1} \frac{H_\textrm{BAH} ^ {S=1}}{J} = \sum_{i=0} \boldsymbol{S}_{2i + 1} \cdot \boldsymbol{S}_{2i + 2}.
\end{align}
The ground state of this Hamiltonian is simply a valence bond in the bulk of the chain, with the first and last spins being left uncoupled to the rest of the chain, resulting in a 9-fold degenerate ground state. In this case we find that the ground state changes when going from $d=0$ to $|d| = 1$, which is the fingerprint of a phase transition. Thus, we conclude that in this limit, there is a critical dimerization value at which the phase transition takes place. This is consistent with the colormap of Fig. \ref{fig:dc_colormap}, where one finds that for $\beta_2 = 0$ and $\beta_1 \rightarrow 1/3$ $|d_c| \rightarrow 0.18$.

\section{Details of tight-binding and CI-CAS}
\label{Extra CAS and TB}

\subsection{Tight-binding model}
To model the single particle properties of nanographenes, the following tight-binding Hamiltonian in the $\pi$-orbital basis \cite{ortiz2019exchange,jacob22} is used:
\begin{align}
    H_\textrm{TB} = t \sum_{\langle i,j\rangle, \sigma}  c^\dagger_{i,\sigma} c_{j,\sigma} +  t_3 \sum_{\langle\langle\langle i,j\rangle\rangle\rangle, \sigma}  c^\dagger_{i,\sigma} c_{j,\sigma} \label{eq:HTB}
\end{align}
where $t$ and $t_3$ refer to the first and third neighbor hopping, respectively, and $c^{(\dagger)}_{i,\sigma}$ is the annihilation (creation) operator of an electron with spin $\sigma = \{\uparrow, \downarrow \}$ in the $\pi$-orbital located at site $i$. We consider $t = -2.7$eV and $t_3 = t/10$ \cite{tran2017third} throughout the text. Including third neighbor hopping is crucial to allow for kinetic intermolecular hybridization between the zero modes of the building blocks \cite{jacob22}, since these are localized in the majority sublattice and the monomers are covalently linked through the minority sublattice. Including second neighbor hopping would break the bipartite nature of the model and lead to particle-hole asymmetry, which is known to be small; based on this, we do not include this parameter in equation (\ref{eq:HTB}). Electron-electron interactions are taken into account by adding an on-site Hubbard repulsion:
\begin{align}
    H_\textrm{Hubb} = U \sum_i n_{i,\downarrow} n_{i,\uparrow} \label{eq: Hubb}
\end{align}
with $U$ the Hubbard repulsion parameter and $n_{i,\sigma} = c^\dagger_{i,\sigma} c_{i,\sigma}$; $U=|t|$ is considered throughout the text. The role of this term is to generate an energy penalty, $U$, for the double occupation of a given $\pi$-orbital

\subsection{CI-CAS}
Due to the exponential growth of the Hilbert space as the system size increases, this Hamiltonian can only be solved exactly for rather small systems (roughly up to 20 sites). The molecules we consider in this work are too large to exactly diagonalize the Hubbard model, so, in order to solve it, approximations need to be introduced.
The Configuration Interaction in the Complete Active Space (CI-CAS) \cite{cusinato2018electronic} approximation is a commonly employed technique to obtain approximate solutions to the Hubbard model in systems where exact diagonalization is out of computational reach. The idea behind this approximation is simple: first, we express the Hubbard Hamiltonian in the basis of the single particle solutions\cite{ortiz2019exchange} ($U=0$); then, we restrict the Hilbert space by considering only a subset of $N_O$ single particle states, which define the active space; finally we account for all possible configurations of $N_e$ electrons distributed over those $N_O$ orbitals, and keep the occupation of the states outside of the active space fixed, with the states below (above) the active space being doubly occupied (empty). For systems at half-filling, which is the case we consider, we have $N_O = N_e$, and we label the approximation as CAS($N_e, N_O$).

\subsection{Results for alternative dimers}
\begin{figure*}
    \centering
    \includegraphics[width = \linewidth]{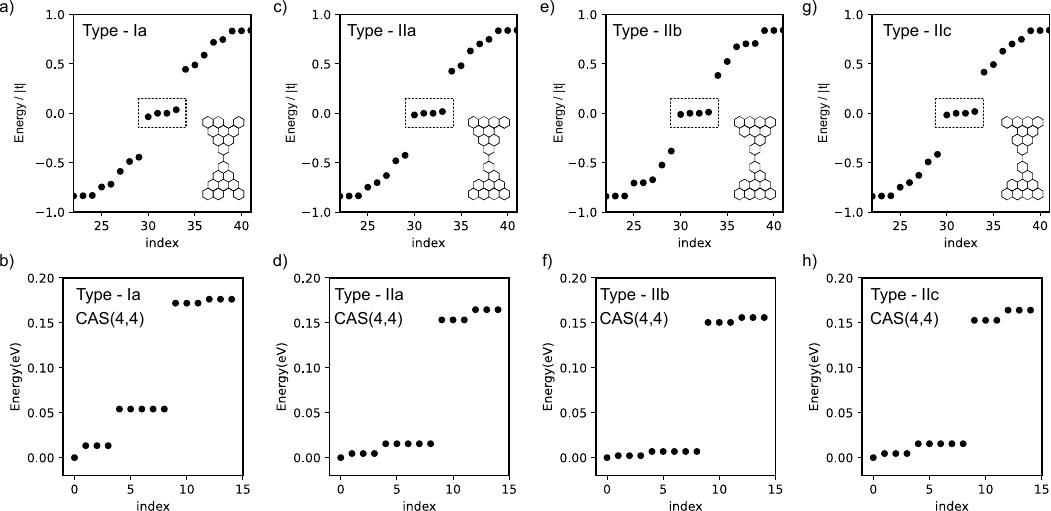}
    \caption{a), c), e) and g) show the single particle spectrum obtained from the tight-binding model given in the main text for the alternative passivated [4]-triangulene dimers. First neighbor hopping was taken as $t = -2.7$ eV, while for third neighbor hopping $t_3 = t/10$. b), d), f) and h) show the solutions of the Hubbard model of the dimers directly above each panel in the CI-CAS approximation, using $U = |t|$.}
    \label{fig:Alternative dimers}
\end{figure*}

Now, we show the single particle spectrum and the CI-CAS solutions of the Hubbard model for the alternative passivated [4]-triangulene dimers shown in the main text. The results are summarized in Fig. \ref{fig:Alternative dimers}, and the bilinear and biquadratic exchange values given in the main text were obtained from these results.
Just like in the main text, we find a larger splitting in the zero modes of the Type-Ia dimer than its Type-II counterparts, which is also reflected in the bigger exchange in the CAS calculation. This is directly seen from the larger energy splitting between the states in the spin sector (which is split from the ionic sector by a significant energy gap, justifying the use of the CI-CAS approximation). For all dimers, only the 4 single particle states closest to zero energy were considered due to the existence of a large gap between these and the remaining molecular orbitals.

\subsection{\bluecolor{Convergence with active space size and $U$}}
\bluecolor{
In this section we shall discuss how the results given in the main text depend on the number of orbitals included in the active space, as well as their dependence on the Hubbard repulsion $U$.
}

\bluecolor{
In Fig. \ref{fig:CAS_vs_nO} we show the solutions of the Hubbard model for all the dimers considered in the manuscript as a function of the number of orbitals ($N_O$) included in the calculation. In this figure we find, in general, a reasonable agreement between the results obtained from different active spaces, although some features, which we discuss below, stand out.
\begin{figure}
    \centering
    \includegraphics[width=\linewidth]{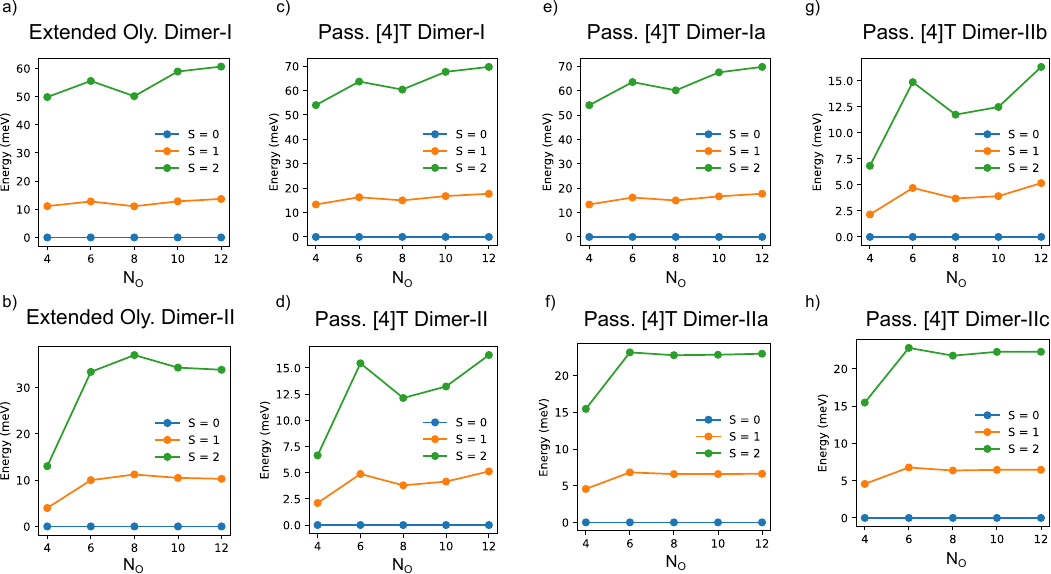}
    \caption{Energies of the first three states obtained from the diagonalization of the Hubbard Hamiltonian in a CAS($N_O$, $N_O$) approximation for the 8 dimers considered in the manuscript. The labels indicate the spin $S$ of each state. The parameters $t = -2.7$eV, $t_3 = t/10$ and $U = |t|$ were used.}
    \label{fig:CAS_vs_nO}
\end{figure}
}

\bluecolor{
To perform a more in-depth analysis, we start by pointing out that the active space should be chosen in such a way that there is significant energy gap between the first state of the active space, and the first state outside of it. As one sees, for example, from the single particle states of Fig. \ref{fig:Alternative dimers}, such a gap is hard to identify when the active space is extended beyond the zero modes. When several states are close in energy, it is possible that their contributions to the final result come with similar magnitudes and opposite signs. Therefore, if one is not careful in the choice of the active space, spurious contributions may be included. In \cite{jacob22}, it was shown how the convergence of results obtained from CAS is often not straightforward.
}

\bluecolor{
Let us now look at some specific cases, starting with panel b), regarding the Type-II extended-Oly. dimer. There we see a sharp increase of the excited states energies when going from CAS(4,4) to CAS(6,6), which is not surprising given the proximity of two orbitals to the zero modes, which justifies the use of the latter approximation in the main text. Further increasing the active space introduces only small corrections in the first two excited states' energies. In panels d) and g), regarding the Type-II and Type-IIb Extended [4]T dimers, we find an energy increase for the excited states when going from CAS(4,4) to CAS(6,6), followed by a decrease when two more orbitals are included in CAS(8,8), highlighting how consecutive states can give contribute symmetrically to the overall result. Even though we show results of CAS(10,10) and CAS(12,12) in all cases, the quality of these approximations is hard to infer since the additional orbitals are part of clusters of nearly degenerate states.
}

\bluecolor{
Based on these considerations, we chose to consider the minimal CAS(4,4) results for all dimers, except for the extended-Oly. Type-II dimer where CAS(6,6) was used instead, since: i) this corresponds to the safest choice of active space; ii) the results are qualitatively similar; iii) this already gives results consistent with those independently obtained from DFT.
}

\bluecolor{
In Fig. \ref{fig:CAS_vs_U} we show how the CAS results depend on the Hubbard repulsion $U$. 
\begin{figure}
    \centering
    \includegraphics[width=\linewidth]{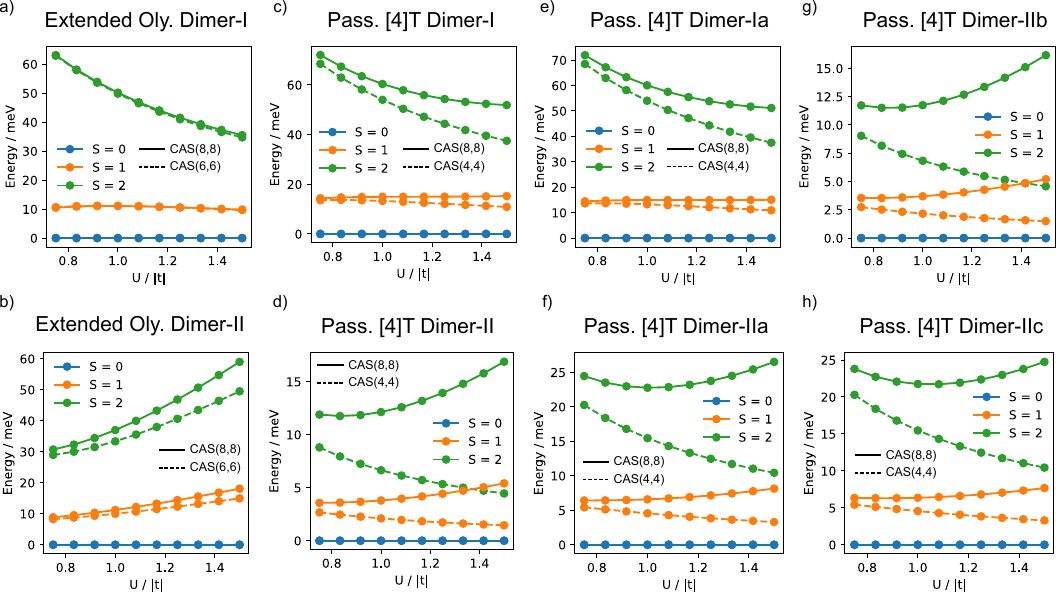}
    \caption{Energies of the first three states obtained from the diagonalization of the Hubbard Hamiltonian in the indicated CAS approximations for the 8 dimers considered in the manuscript as a function of the Hubbard repulsion $U$. The labels indicate the spin $S$ of each state. The parameters $t = -2.7$eV, $t_3 = t/10$ were used.}
    \label{fig:CAS_vs_U}
\end{figure}
When only the zero modes are included in the active space, the only active exchange mechanism is kinetic exchange \cite{jacob22}, which leads to a $1/U$ behavior. Including more orbitals in the active space allows for Coulomb driven exchange to be present, which leads to a $U^2$ asymptotic behavior \cite{jacob22} at large $U$ values. In the physically relevant region for $U$, i.e. in when $U \sim |t|$, the CAS results do not change abruptly under small variations of this parameters.
}

\subsection{\bluecolor{Trimers and additional spin interactions}
\label{sec:additional}
}
\begin{figure}
    \centering
    \includegraphics[width=\linewidth]{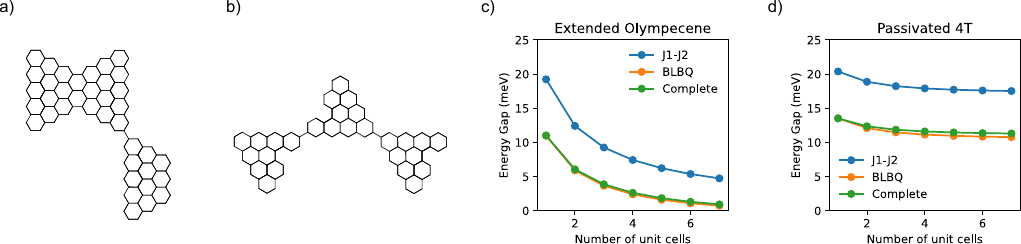}
    \caption{a) Extended-Olympicene trimer; b) Passivated [4]-triangulene trimer; c) Evolution of the first excited state gap as a function of the spin chain length for three spin models using the exchange parameters obtained from fitting the spin model to the results obtained from CAS for the Extended-Olympicene trimer; d) Same as panel c) but using the parameters obtained from the fit to the passivated [4]-triangulene trimer.}
    \label{fig:spin model comparison}
\end{figure}

\bluecolor{
In the spin model of equation \ref{Eq: S=1 BAH} only bilinear and biquadratic interactions were accounted for. As we pointed out in the main text, in a previous work \cite{henriques2023anatomy} other spin interactions were derived from the same order of perturbation theory as the biquadratic terms. These additional terms read:
\begin{align}
    H^\textrm{extra} = J_3 \sum_i \boldsymbol{S}_i \cdot \boldsymbol{S}_{i+2} + B_3 \sum_i (\boldsymbol{S}_i \cdot \boldsymbol{S}_{i+1}) (\boldsymbol{S}_{i+1} \cdot \boldsymbol{S}_{i+2})
\end{align}
To determine the values these new interactions we have to go beyond the dimers we have considered so far, and study trimers instead, since at least three units are needed to compute $J_3$ and $B_3$. These trimers are depicted in Fig. \ref{fig:spin model comparison}a) and b).
}

\bluecolor{
To obtain $J_3$ and $B_3$ we follow a similar procedure to the one described in the main text to find the bilinear and biquadratic exchanges. First, we solve the Hubbard model for these new systems in a CAS approximation. Then, we fit the eigenvalues to the energy levels obtained from CAS to find the optimal parameters of the spin-model. For the extended-Oly. trimer we perform CAS(8,8) to include the 6 zero modes and two nearby orbitals (similar to what was found in the Type-II dimers); for the passivated [4]-triangulene trimer CAS(6,6) is used instead. Following this procedure, the following parameters for the extended-Oly. are found: $J_1 = 19.23$meV, $J_2 = 11.29$meV, $B_1 = 2.74$meV, $B_2 = 0.57$meV, $B_3 = - 0.15$meV and  $J_3 = 0.36$meV. From the passivated [4]-triangulene trimers we find $J_1 = 20.37$meV, $J_2 = 2.21$meV, $B_1 =2.28$meV, $B_2 = 0.17$meV, $B_3 = 0.09$meV, $J_3 = 0.09$meV. From both of these fitting procedures, we find that the new interactions ($B_3$ and $J_3$) are smaller than the smallest of the biquadratic terms by a factor of at least two, and are much smaller than the others exchanges (whose value remain consistent with the ones presented in Table I when extracted from the dimers).
}

\bluecolor{
Now, to understand the impact of these terms in the properties of the spin system, we study how the gap of the spin Hamiltonian evolves with chain length for three different models: i) the $J_1$-$J_2$ model; ii) the BLBQ Hamiltonian of equation \ref{Eq: S=1 BAH}; and iii) the complete model where $H^\textrm{extra}$ is also included. In all cases, we consider the first exchange in the chain to be the largest one. The results obtained with the parameters of the extended-Oly. and passivated [4]-triangulene trimers are depicted in Fig. \ref{fig:spin model comparison}c) and d). The first thing we notice is that, for both types of system, the topological phase remains the same for the three models, i.e. for the extended-Oly. case we always have a system in the Haldane phase with a gap that exponentially approaches zero with system size, and for the passivated [4]-triangulene case we have a gapped system. Crucially, we find that when biquadratic terms are included in the $J_1-J_2$ Hamiltonian, a big change is visible in the evolution of the gap. On the other hand, we find that further including $J_3$ and $B_3$ introduces minor corrections only. Based on all of the above, we conclude that these additional terms can safely be neglected, and equation \ref{Eq: S=1 BAH} is enough to describe these nanographene chains.
}
\section{DFT Calculations}
\label{Extra DFT Calc}

Density Functional Theory (DFT) results where obtained using the {\sc QUANTUM ESPRESSO} package \cite{QE-2017, QE-2009, doi:10.1063/5.0005082}. We used norm-conserving pseudopotentials\cite{PhysRevLett.43.1494, PhysRevB.43.1993, van_Setten_2018, Hamann2013, garrity2014pseudopotentials, doi:10.1126/science.aad3000}. We used a plane-wave energy cutoff of 70 Ry for the wavefunctions and 280 Ry for the charge density. The Brillouin zone integrals were performed in a 1$\times$1$\times$1 k-point grid with a Gaussian smearing of 0.001 Ry. \bluecolor{The nanographene dimer and monomer structures include a converged 9 \AA\space vacuum for every axis to avoid inter-molecular interactions  \bluecolor{and H atoms bonded to all edge C atoms. All structures where relaxed optimizing the ionic positions while keeping the unit cell parameters fixed in order to preserve the converged vacuum. The ionic relaxations where performed using BFGS quasi-newton algorithm where the convergence threshold on forces was 10$^{-3}$ a.u.}}

\begin{figure*}
    \centering
    \includegraphics[width=0.8\textwidth]{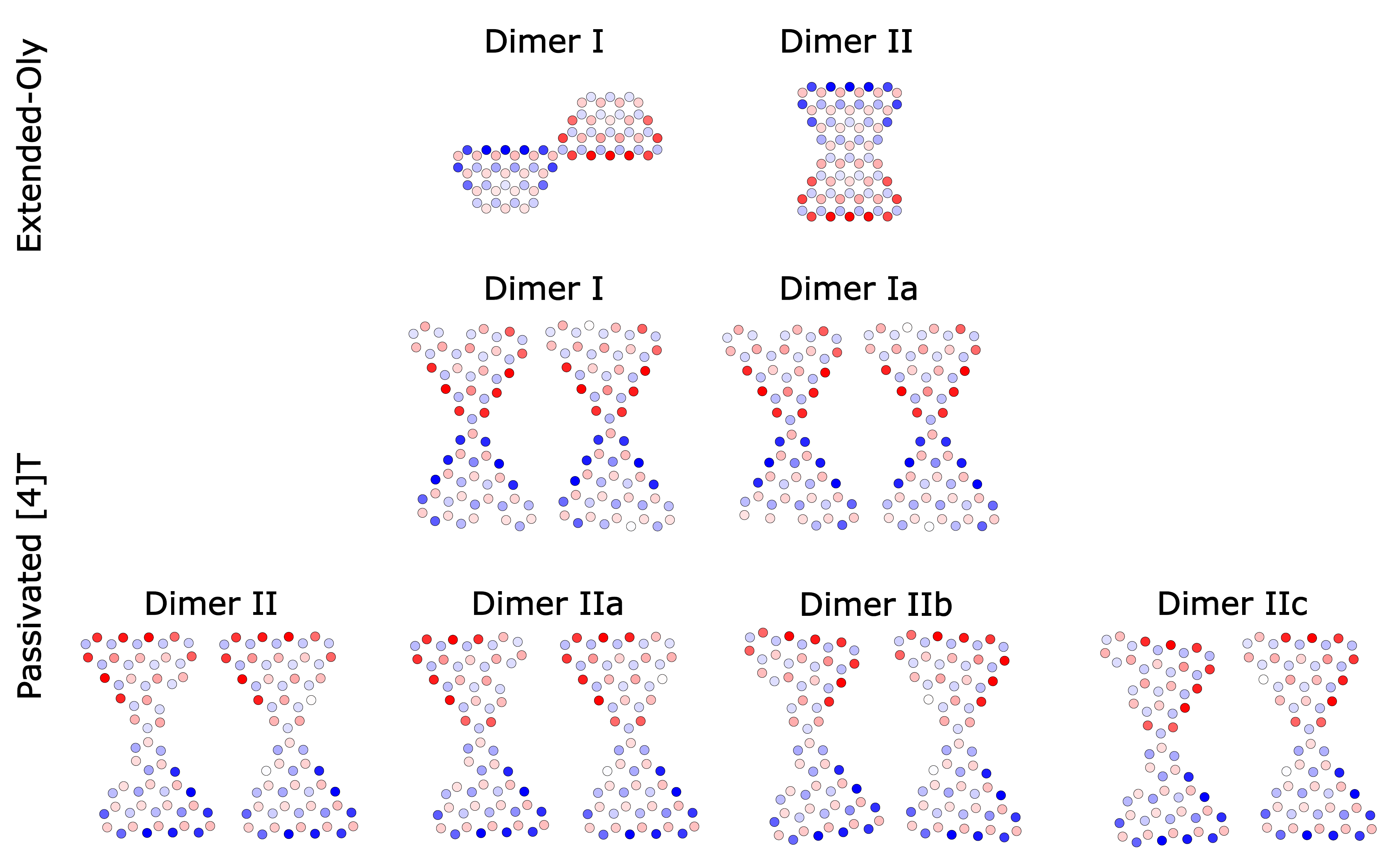}
    \caption{DFT magnetic profiles \bluecolor{of the antiferromagnetic solution} for the C atoms of the different analyzed dimers. First row (Extended-Olympicene) depicts the two different considered dimers for the olympicene molecule. 
    Second row (Passivated [4]T) depicts the different passivation combinations for the [4]T dimers. For each dimer we have compared the C-removal passivation (left) and the H-passivation (right).}
    \label{fig:dft_results}
\end{figure*}

Passivation of the [4]-triangulene dimers was done through two different approaches. The first method consisted on removing a C atom at specific sites at each [4]-triangulene \bluecolor{with subsequent sigma-bond passivation using extra H atoms in the newly formed edge C atoms}. The second method consisted on adding an extra H bond to the C atoms which we intended to passivate. Results of the computed DFT magnetic moments are shown in Fig. \ref{fig:dft_results} for the Extended-Olympicene and the passivated [4]-triangulene dimers. We also computed the \bluecolor{bilinear} exchange values for the different passivation methods. Using a quantum spin model where the energy difference of the AFM and FM solutions correspond to the singlet and quintuplet respectively:

\begin{align}
    E_{FM}-E_{AFM} = 3J.
\end{align}
obtaining an excellent degree of agreement between both approaches. In our convention, positive values of $J$ imply that the AFM solution is favored. The computed \bluecolor{bilinear} exchange values are presented in Table S1.

\begin{table}[]
\begin{tabular}{|l|l|l|l|l|}
\hline
\multicolumn{1}{|c|}{\multirow{2}{*}{Monomer}} & \multicolumn{1}{c|}{\multirow{2}{*}{Dimer}} & \multicolumn{1}{c|}{C-removal}       & \multicolumn{1}{c|}{H-passivation}                                        \\ \cline{3-4} 
\multicolumn{1}{|c|}{}                         & \multicolumn{1}{c|}{}                       & $J$ (meV) & $J$ (meV)  \\ \hline

\multirow{6}{*}{Pass. 4{[}T{]}}                & I                                           & 16.49                           & 16.86 \\ \cline{3-4} 
                                               & II                                          & 2.51                            & 2.65  \\ \cline{3-4} 
                                               & Ia                                          & 16.51                           & 16.93 \\ \cline{3-4} 
                                               & IIa                                         & 5.03                            & 5.16  \\ \cline{3-4} 
                                               & IIb                                         & 2.87                            & 2.78  \\ \cline{3-4} 
                                               & IIc                                         & 5.05                            & 5.10  \\ \hline
\end{tabular}

\label{tab:dft_exchange_supl}
\caption{Values of the \bluecolor{bilinear} exchange obtained using DFT calculations with two different types of passivation in the 4[T] dimers. The labels of the different passivation options are listed in the second column (Dimers). The numbers in the third column (C-removal) correspond to the \bluecolor{bilinear} exchange of the dimers that where passivated by removing a C atom. The numbers in the fourth column (H-passivation) correspond to the \bluecolor{bilinear} exchange of the dimers that where passivated by putting an H atom on top of the site we wanted to passivate.}
\end{table}

\section{Modelling $dI/dV$ spectroscopy}
\label{dIdV details}
To model the $dI/dV$ spectra, we assume a spin chain that is coupled to two reservoirs: the STM tip and the substrate. We consider two types of electron scattering that can produce a spin flip in a given site of the chain: electrons that tunnel from the tip to the sample, exciting the spin chain in the process; and scattering between the substrate electrons. The $dI/dV$ is obtained by first computing the current $I$ with scattering theory including corrections up to the third order \cite{ternes2015spin}, followed by the derivative with respect to the bias $V$ (which we define as the difference between the chemical potential of the reservoirs). Up to the second order, the $dI/dV$ is composed of excitations step, centered at the energy difference between the ground state and the excited state, and whose height is determined by the spin spectral weight. The spin spectral weight associated to a transition from a state $M$ to a state $M'$ due to a perturbation on spin site $i$ is given by:
\begin{align}
    W_{M,M'}(i) = p_M \sum_{a=x,y,z} |\langle M'|S_a(i)|M\rangle|^2,
\end{align}
where $p_M$ is the thermal occupation of state $M$. The matrix element of the spin operator between the initial ($M$) and final ($M'$) states, enforces the spin selection rule $\Delta S = 0,\pm1$. The third-order corrections account for processes mediated by intermediate states; the spin selection rule is enforced for all excitations, but energy conservation is only required between the initial and final states. The third-order correction is responsible for the introduction of logarithmic resonances whose main effects in the simulation are: (1) overshooting features superimposed on the thermally broadened excitations steps; (2) a Kondo peaks at zero bias for degenerate ground state. In our simulations we consider Kondo coupling of $J_K \rho = 0.15$ and an effective temperature of $T_\textrm{eff} = 5$K. This $T_\textrm{eff}$ is an effective parameter which is used to describe the experimental linewidth, and accounts for, in an empirical manner, all the broadening channels (i.e. not only the thermally induced linewidth). The exact diagonalization of the spin Hamiltonian is performed using the QuSpin package \cite{weinberg_quspin_2017}.

\end{appendix}






%

\end{document}